\documentclass[a4paper,journal]{IEEEtran}
\usepackage{graphicx,amsmath,bm}
\usepackage{graphicx}
\usepackage{latexsym}
\usepackage{graphicx}
\usepackage{rotating}
\usepackage{amsmath}
\usepackage{amssymb}
\usepackage{url}
\usepackage{geometry}
\usepackage{pdflscape}
\usepackage{balance}
\usepackage{marvosym}
\usepackage{algorithmic}
\usepackage{algorithm}
\usepackage{multirow}
\usepackage{multicol}
\usepackage{lipsum}
\usepackage{psfrag}
\usepackage{microtype}
\usepackage{vwcol} 
\usepackage{caption}
\usepackage[table]{xcolor}
\usepackage{color}
\usepackage{pifont}
\newcommand{\cmark}{\ding{51}}

\usepackage[htt]{hyphenat}
\usepackage{capt-of}
\usepackage[inline]{enumitem}
\usepackage[table]{xcolor}
\newcommand{\envelope}{(\raisebox{-.5pt}{\scalebox{1.45}{\Letter}}\kern-1.7pt)}
\usepackage{times}
\usepackage{pdflscape}
\usepackage{lscape}
\begin{document}

%%paper title
%%For line breaks, \\ can be used within title 
\title{QoS--aware Mesh based Multicast Routing Protocols in Ad-Hoc Networks: Concepts and Challenges}

%%author names are separated by comma (,) use \and before the last author name \textsuperscript{number} is used for affiliation
%%use a * along with the number separated by comma for the  author for correspondence

%\author{Gaurav Singal\textsuperscript{1,*}, Vijay Laxmi\textsuperscript{1}, Manoj S. Gaur\textsuperscript{1}, D Vijay Rao\textsuperscript{2} \and Riti Kushwaha\textsuperscript{1}}
%\affilOne{\textsuperscript{1} Department of Computer Science and Engineering, MNIT Jaipur\\Email:{gauravsingal789@gmail.com, gaurms@mnit.ac.in, vlaxmi@mnit.ac.in, riti.kushwaha07@gmail.com\\}}
%\affilTwo{\textsuperscript{2} Institute for Systems Studies and Analyses, DRDRO, Delhi, doctor.rao.cs@gmail.com}
%\corres Department of Computer Science \& Engineering, Malaviya National Institute of Technology, Jaipur - 302017, India. E-mail: gauravsingal789@gmail.com
%%escape two column mode for title, affiliation and abstract by giving \twocolumn command as shown

\author{Gaurav Singal,%~\IEEEmembership{Member,~IEEE,}
	~Vijay Laxmi,%~\IEEEmembership{Fellow,~OSA,}
	~Manoj S Gaur,%~\IEEEmembership{Fellow,~OSA,}
	~D Vijay Rao,%~\IEEEmembership{Fellow,~OSA,}
	~and~Riti Kushwaha,%~\IEEEmembership{Life~Fellow,~IEEE}% <-this % stops a space
	\thanks{G. Singal is with the Department of Computer Engineering, Bennett University, India,
		201308 e-mail: gaurav.singal@bennett.edu.in}% <-this % stops a space
	\thanks{M. Gaur, V. Laxmi and R. Kushwaha are with Malaviya National Institue of Technology Jaipur, India.}% <-this % stops a space
	\thanks{D V. Rao is with Defence Research and Development Organization Delhi, India}% <-this % stops a space
	}

\markboth{QoS--aware Mesh based Multicast Routing Protocols in Ad-Hoc Networks}%
{Shell \MakeLowercase{\textit{et al.}}: Bare Demo of IEEEtran.cls for IEEE Journals}

%\twocolumn[
\maketitle
\begin{abstract}
Multicast communication plays a crucial role in Mobile Adhoc Networks~(MANETs). MANETs provide low cost, self configuring devices for multimedia data communication in military battlefield scenarios, disaster and public safety networks~(PSN). Multicast communication improves the network performance in terms of bandwidth consumption, battery power and routing overhead as compared to unicast for same volume of data communication. In recent past, a number of multicast routing protocols~(MRPs) have been proposed that tried to resolve issues and challenges in MRP. Multicast based group communication demands dynamic construction of efficient and reliable route for multimedia data communication during high node mobility, contention, routing and channel overhead. This paper gives an insight into the merits and demerits of the currently known research techniques and provides a better environment to make reliable MRP. It presents a ample study of various Quality of Service~(QoS) techniques and existing enhancement in mesh based MRPs. Mesh topology based MRPs are classified according to their enhancement in routing mechanism and QoS modification on On-Demand Multicast Routing Protocol~(ODMRP) protocol to improve performance metrics. This paper covers the most recent, robust and reliable QoS and Mesh based MRPs, classified based on their operational features, with their advantages and limitations, and provides comparison of their performance parameters.
\end{abstract}
\begin{IEEEkeywords}
	MANETs, Multicast Routing Protocol, Mesh Based, QoS
\end{IEEEkeywords}
%\keywords{MANETs, Multicast Routing Protocol, Mesh Based, QoS}
%]
%%manuscript information received, revised and accepted dates
%%
%\msinfo{15 September 2016}{15 September 2016}{15 September 2016}
%%insert keywords separated by comma

%%close the twocolumn escape here

%%include \corres to print the footnote for correspondence
%%include \volnum{number} for the volume number in the header
%%include \monthyear{month year} for month and year of publication in the header
%%include \pagerange{num--num} page range of article in header
%%include \doi{number}for the DOI number in the header

%\corres
%\volnum{123}
%\issuenum{4}
%\monthyear{January 2016}
%\pgrange{2333--2337}
%\doinum{12.3456/s78910-011-012-3}
%

\IEEEpeerreviewmaketitle

%%The running head information

%\markboth{Gaurav Singal, et. al.}{Mesh Based MRP in MANETs}
%\input{./sections/introduction.tex}
\section{Introduction}
\label{intro}
%Mobile Ad hoc Networks~(MANETs) came into existence to handle the undesirable situations occurred due to disasters such as earthquakes, floods, fires and terrorist attacks. 

Mobile Ad hoc Networks~(MANETs) came into existence with an aim to handle the undesirable situations that may occur due to disasters such as earthquakes, floods and fires or due to human activities like terrorist attacks, military operations and so on. MANET is a type of wireless communication network which do not have any stable infrastructure and central administrative control. In past few years, MANETs have been deployed for other purposes like audio/video conferencing, emergency rescue operations, traffic control and online lectures~\cite{Gaurav:2016}.

\par These networks possess some excellent features such as fast deployment, flexibility, robustness, mobility support and highly dynamic network topology (fading, shadowing, network partition)~\cite{Badarneh:2009}. In MANET, node can communicate with relay (intermediate) nodes if communicating host nodes are not in its range (multi-hop routing).
%Mobile Ad hoc Networks~(MANETs) represents a type of wireless communnication networks. These networks do not have any stable infrastructure and center administrative control. MANET's comes into existance to address the undesirable situation occured due to calamity such as natural crises like earth quakes, floods, fires and terrorist attacks. Now, the idea of MANET's has been used to enhance applications such as Audio-Video Conferencing, emergency rescue operations, traffic control, online lectures. 

%MANET  is a group of wireless mobile nodes or router. The routers are able to walk arbitrary. MANET is self-organized and can be  
\par MANET is a group of wireless mobile nodes that may act as host as well as router and are able to move arbitrarily. MANET is a self-organized network that can be deployed anywhere, at any time to support particular conditions. In contrast to MANETs, infrastructure-dependent wireless networks are more reliable and provide Quality of Services~(QoS) assurance. The unreliability in MANETs occur due to limited battery power, limited bandwidth (channel capacity), heterogeneity, high routing overhead and unpredictable node mobility. Bandwidth, Delay, Signal Strength and other metrics are used for QoS assurance in multicast group communication for both data and real-time traffic.

%These networks possess some excellent features such as fast deployment, flexibility, robustness, mobility support and highly dynamic network topology (fading, shadowing, network partition)~\cite{Badarneh:2009}. In MANET, node can communicate with relay (intermediate) nodes if communicating host nodes are not in its range (multi-hop routing).

\par In recent years, multicasting has been greatly appreciated in any type of group communication like audio/video conferencing, video lectures. Multicast Routing Protocol~(MRP) communicates datagram to a group of destinations recognized by single multicast address at single transmission time. Multicast transmission helps to improve node energy, congestion on channel capacity, time and resource utilization as compared to Unicast Routing Protocol~(URP), in case of transmission of datagram to a group of destinations. Multicasting in MANETs is more complex than wired networks, in terms of node energy and bandwidth. High mobility, low channel capacity and battery issues attracted attention of many researchers towards multicast routing protocols to build robust, reliable and scalable networks.
\par In recent past, numerous surveys have been published on Multicast Routing Protocol in MANETs. In~\cite{Junhai:2009}, the authors have placed routing protocols in two broad categories: Multicast Routing Protocol based on application dependence and application in-dependence. In~\cite{Badarneh:2009}, authors classified the existing MRPs into three categories according to their layer of operation namely, the Network layer, the Application layer and the MAC layer. In~\cite{RajManvi:2012}, authors classified MRPs based on routing protocol mechanisms and focused on reliable and QoS aware MRP.

%\par Explain all parameter...like congestion, link stability, etc.
\par Multicast routing protocols have been improved by the researchers consistently on the basis of various evaluation metrics like quick route recovery, reliability, improved QoS~(less energy consumption, reduce channel capacity utilization), less congestion(interference), improved Packet Delivery Ratio~(PDR) and end-to-end delay, network life time and last but not the least, security. Group communication faces many challenges and issues such as resource management, synchronization, power management and routing management~\cite{RajManvi:2012,Gauravmclspm:2016}. Real time applications require reliable and stable communication among multicast group members.
%People have proposed various multicast routing mechanism to improve routing performance in MANETs under QoS based, routing based or Security based.
\par None of the authors has given clear picture of mesh based MRPs and their proposed modifications. In this paper, we have explored mesh based MRPs and classified them on the basis of modifications in routing mechanism and QoS metrics adaptation. We have explained some issues and challenges in designing MRPs such as energy efficiency, reliability, security and QoS aware multicasting. At last, we have presented a taxonomy of mesh based MRPs on the basis of their techniques, features, modification components and improvement parameters. We have also given a comparative chart between proposed enhancements in ODMRP protocol.
%\begin{enumerate}
%\item PCs are more enterprise friendly as compared to mobile platforms in terms of using %\item Malware authors do not intentionally aim at Windows because they think it is more %\end{enumerate}
\par Rest of the paper is organized as follows: Section~\ref{Issue and challenges} discusses the design issues and challenges in multicast routing protocol. Current state of multicast routing protocols has been described in Section~\ref{taxonomy}. We further described the proposed modifications in mesh based protocol in Section~\ref{mesh_based}. Section~\ref{linkstability} elaborates the multiple mechanisms and requirement of estimating link stability. Finally, the future directions and concluding remarks are given in Section~\ref{future} and Section~\ref{conclude} respectively.
%%%%%%%%%%%%%%%%%%%%%%%%%%%%%%%%%%%%%%%%%%%%%%%%%%%%%%%%%%%%%%%%%%%%%%%%%%%%%%%%%%%%%%%%%%%%
%\input{./sections/Taxonomy.tex}
\section{Taxonomy of Multicast Routing Protocol}
\label{taxonomy}
Wired network is still the simplest way to use Internet. Now-a-days, mobile and portable devices demand internet connectivity at home, at work and on walk. Wireless networks are categorized as infrastructure and infrastructure less networks. Cellular network is an example of infrastructure network, with high set-up cost and time. Adhoc network is an example of infrastructure less network with cost-effectiveness and less set-up time.
\begin{figure*}[!htbp]	
	\begin{center}
		\includegraphics[scale=0.4]{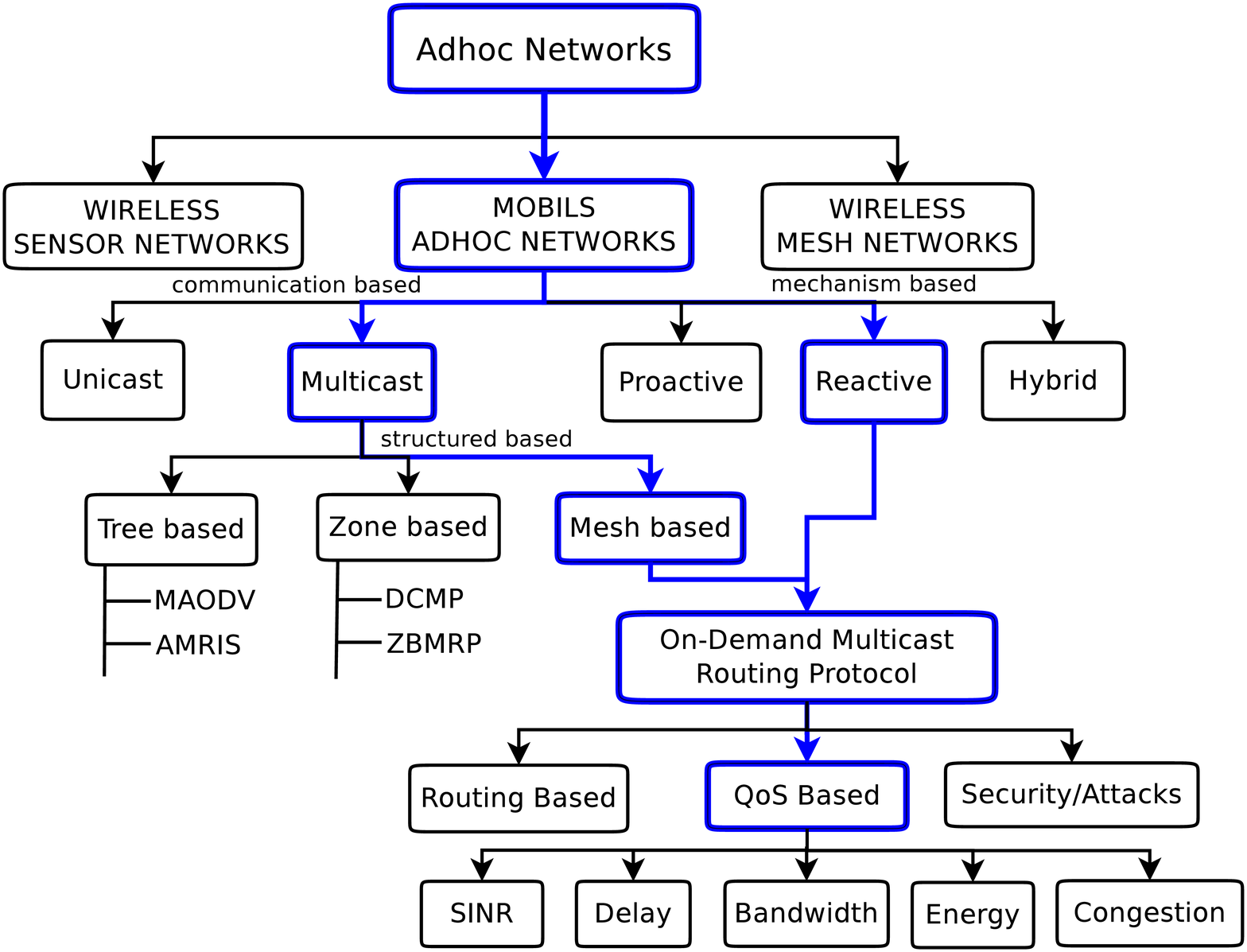} 
		\caption{\small{Taxonomy of Routing Mechanism in MANETs}}
		\label{fig:taxonomy}
	\end{center}
\end{figure*}
\par Adhoc means "for the purpose", self-organizing network architecture. There is no requirement of base station. Adhoc networks are further classified as Mobile Adhoc Networks~(MANETs), Vehicular Adhoc Networks~(VANETs), Wireless Sensor Network~(WSN), Wireless Mesh Network~(WMN). Here, we focus on routing protocols for MANETs.
\par Routing protocols for MANETs can be categorized on the basis of mechanism as reactive~(routes are created on demand), proactive~(pre-determined routes are stored in routing tables and are periodically updated) and hybrid~(some nodes have predefined and some have on-demand). In terms of number of destinations, that a protocol can transmit data for a given source, routing can be Unicast (only one destination supported) or Multicast (for group of destinations). Figure~\ref{fig:taxonomy} presents an overall picture of routing mechanisms in MANETs.
\begin{figure*}[!htbp]
	\begin{center}
		\includegraphics[scale=0.4]{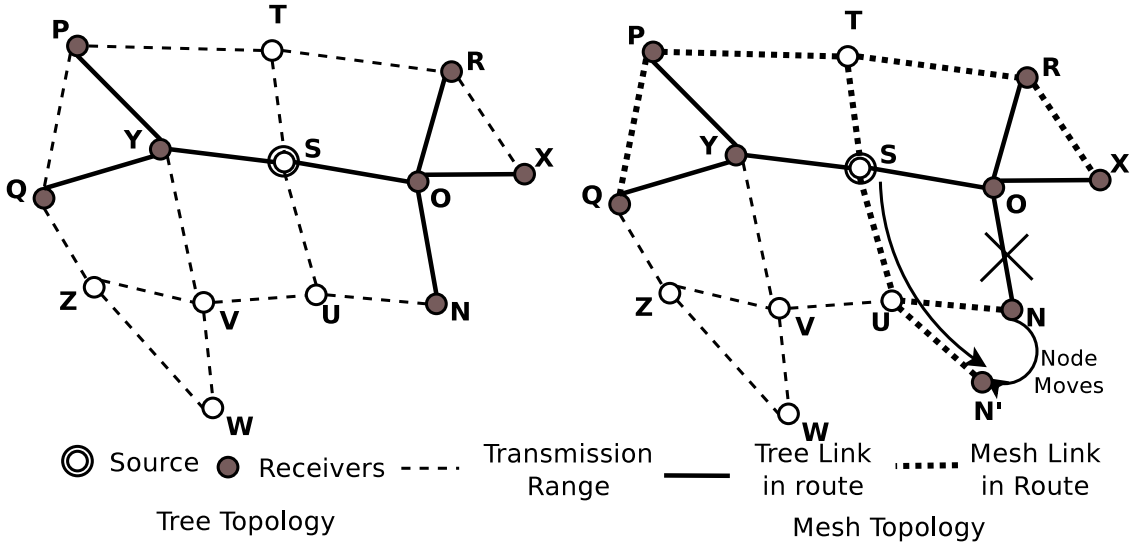} 
		\caption{\small{Difference between Tree and Mesh Topology}}
		\label{fig:meshtree}
	\end{center}
\end{figure*}
\par Numerous unicast routing protocols~\cite{Marina:2005} supporting Quality of Service~(QoS) have been explored. These protocols provide a stable path from a source to single destination. We need to explore Multicast Routing Protocols~(MRP)~\cite{Luo:2009,RajManvi:2012,Gauravwiley:2016} addressing the limitations of unicast routing protocols. MRPs can be classified on the basis of their routing structure as (1)~tree-based, (2)~zone-based, (3)~mesh-based, (4)~hybrid. Tree based MRP is very efficient in routing in network and provides better packet delivery ratio as compared to other protocols, but there is excessive reconfiguration overhead in case of re-routing. In tree based MRP, there are many protocols such as Multicast Ad-hoc On demand Distance Vector routing~(MAODV)~\cite{Royer:1999}, Ad-hoc Multicast Routing protocol utilizing increasing Id numberS~(AMRIS)~\cite{Wu:1999}, Ad-hoc Multicast Routing protocol~(AMRoute)~\cite{Xie:2002}, etc.

For limited reconfiguration and rebuilding caused by redundancy of packets, mesh based MRP is better than others~\cite{RajManvi:2012}\@. In mesh based MRP, more than one path exists between pair of source and destinations. There are many protocols such as On-Demand Multicast Routing Protocol~(ODMRP)~\cite{Sung:1999}, Enhanced On-Demand Multicast Routing Protocol~(EODMRP)~\cite{Oh:2008}, Forwarding Group Multicast Protocol~(FGMP)~\cite{Chiang:1998} and Team Oriented Multicast~(TOM) Protocol~\cite{Yi:2003} that create mesh structure after route construction.
Zone based MRP forms a cluster of source, receiver and intermediate nodes in routes. Selection of zone leaders on the basis of first announcement for better and robust decisions. There are many protocols that provide zone based structure for transmission such as Dynamic Core based Multicast routing Protocol~(DCMP)~\cite{Das:2002}, Cluster Based Stable multicast Routing Protocol~(CBSRP)~\cite{Xu:2008}.

\par To better understand the topological difference between tree and mesh topologies, we have considered a scenario in which one sender and five receivers are there as shown in Figure~\ref{fig:meshtree}. The major difference in between tree and mesh topologies is the number of alternative paths provided between two nodes. The former provides only one path while the later provides multiple paths for single destination. Due to single path option, tree topologies are not suitable for applications in which link failures and node movement exist.
\par For example, source $S$ is transmitting data to multiple receivers. If link between $O$ and $N$ fails or node $N$ moves to $N'$, then tree topology cannot continue transmission. But mesh topology provides another route to transmit the data due to multiple or alternate paths available between source and destinations.
%\par In figure~\ref{fig:meshtree}, we are presenting difference between tree and mesh topology. We have take a scenario considering one sender and five receiver. Firsty, We have applied tree based MRP for route construction to get single path for corresponding receivers. Then we again aaply mesh based MRP to get redandant paths. In tree topology, N receiver gets packet from O intermediate node. If any case N moves to position N' then tree link is broken and N' not received any packets. In mesh we have multiple paths for single destination. Data packet can receive to N' by O intermediate node, if any link failure~\cite{Sarkar:2007}.
\par In this paper, we have focused on mesh based multicast routing protocols and explained multiple enhancements in ODMRP protocol on the basis of QoS and Routing Modification.
\section{Mesh based Multicast Routing Protocol}
\label{mesh_based}
Mesh topology is robust and reliable for communicating data to the destination in case of node mobility or link failure. It doesn't require reconfiguration of network because there already exist redundant (multiple) paths for every destination.\\
All forwarding group members, multicast group members and links between them form a mesh. The characteristic feature of mesh is that the node doesn't care about upstream node, from which the packet has arrived, and it rebroadcasts non-duplicate packet. If one node lies in the transmission range of other node, then both nodes share a mesh link. So, the mesh structure has more connected links than tree and increases the robustness of multicast group, which is convenient in generous and frequent link breaks for ad-hoc networks~\cite{Yao:2003}.\\
%If two node are directly in their tranmisson range of each other, 
Robustness of ODMRP protocol depends upon number of senders and mobility speed. At low mobility and large number of senders, ODMRP creates redundant routes, some of which may be useless while at high mobility and less number of senders, it offers less redundant routes~\cite{Ghafouri:2009}.

%%%%%%%%%%%%%%%%%%%%%%%%%%%%%%%%%%%%%%%%%%%%%%%%%%%%%%%%%%%%%%%%%%%%%%%%%%%%%%%%%%%%%%%%%%%%%%%%%%%%%

\paragraph{Forwarding Group Multicast Protocol(FGMP) for multi-hop, Mobile Wireless Networks~\cite{Chiang:1998}}
FGMP provides reliability by transmitting data via Forwarding Group and maintains a multicast mesh. In this protocol, both the source and receivers advertise their existence through respective broadcasting packets known as Source Join broadcast (FGMP-SA) Approach and Receiver Join broadcast (FGMP-RA) Approach. When a destination node receives a join request from other node, it updates its own Join table and broadcasts it to other members of group to update their respective table.
\par FGMP reduces overall overhead by limiting flooding within Forwarding Group. FGMP-SA provides better throughput as compared to FGMP-RA in case of less number of senders than receivers in a network. FGMP protocol is not scalable and it does not support high mobility because it gives better results in small network. 
%%%%%%%%%%%%%%%%%%%%%%%%%%%%%%%%%%%%%%%%%%%%%%%%%%%%%%%%%%%%%%%%%%%%%%%%%%%%%%%%%%%%%%%%%%%%%%%%%%%%%%
\paragraph{Core-Assisted Mesh Protocol(CAMP)~\cite{Garcia:1999}}
CAMP has been designed to support multicast routing protocol in mobile adhoc network using a shared mesh structure. In order to limit the control traffic, CAMP uses core node for creating a mesh. To prevent packet replication or looping in the mesh, each node maintains a cache to keep track of recently forwarded packets. The algorithm ensures that all the nodes from reverse shortest path are included in the mesh. Like other core based protocols, it doesn't require whole traffic flow from core nodes.
\par CAMP is based on salient assumption about route information available (proactive) and existence of beaconing protocol. So, it got high overhead because of proactive protocol~\cite{Dhillon:2005}.
%%%%%%%%%%%%%%%%%%%%%%%%%%%%%%%%%%%%%%%%%%%%%%%%%%%%%%%%%%%%%%%%%%%%%%%%%%%%%%%%%%%%%%%%%%%%%%%%%%%%%%
\subsection{On Demand Multicast Routing Protocol}
%\label{odmrp}
\par Sung \textit{et al}~\cite{Sung:1999} have proposed ODMRP~(On-Demand Multicast Routing Protocol), a reactive mesh based adhoc multicast routing protocol that gives reliable routes. This protocol consists of following steps:
\begin{enumerate}
	\item Source sends request of `Join Query' and waits for `Join Reply' from receiver(s). These query packets are sent periodically to whole network. 
	\item On receiving `Join Query' packet, intermediate node rebroadcasts it and sets previous hop address only if received packet has not been seen earlier and discards duplicate packets.
	\item Multicast receiver receives `Join Query' packet from intermediate node(s) and sends `Join Reply' to respective previous hop address. 
	\item On inspecting `Join Reply' packet, an intermediate node checks if the address field matches with its own address. If yes, it creates join table, labels itself as member of forwarding group and forwards the packet to previous hop address.
	\item At last, source receives the join table from intermediate node and selects minimum hop route to forward the data packet. Source also sends acknowledgement to multicast receiver and builds a mesh structure for available route to different destinations.
	\item The periodic transmission is used to refresh the routes and all member tables. 
\end{enumerate}
\begin{figure}[h!]
	\begin{center}
		\includegraphics[scale=0.32]{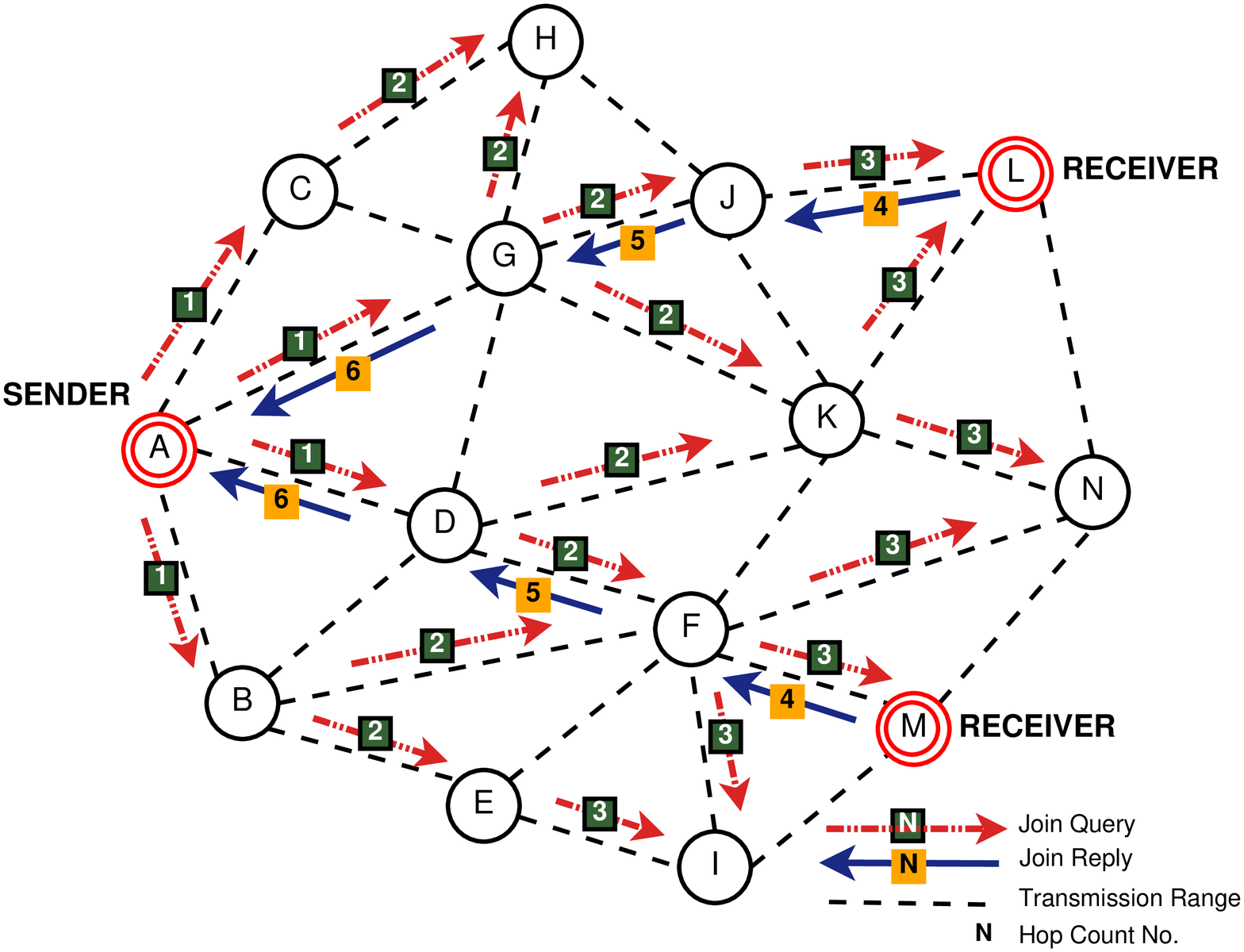} 
		\caption{\small{ODMRP Work Flow}}
		\label{fig:odmrp}
	\end{center}
\end{figure}
\par Work flow of ODMRP protocol is illustrated in Figure~\ref{fig:odmrp}\@. In topology presented here, there is one sender, two receivers (shown inside double ring). In ODMRP, forwarding nodes use the shortest path between multicast group members. Red arrow indicates `JOIN Query' and blue arrow indicates `JOIN Reply'. Weight on an arrow indicates hop count value for respective link. A link marked with both red and blue arrow is part of the path which extends back to source. Information about other possible paths is not discarded and would be used to establish links in case of disconnections induced by mobility. For example, in Figure~\ref{fig:odmrp}, route $A \rightarrow G \rightarrow K \rightarrow L$ is established as soon as $A \rightarrow G \rightarrow J \rightarrow L$ is disrupted because of movement of $J$. As a result, this mesh structure is more resilient over tree-like topology as there is no requirement to reconfigure the entire route, if node's position changes.
\par ODMRP is widely used protocol for group communication in multicast routing protocol due to major advantages of high packet delivery ratio with some limitations like higher control overhead and redundancy of packet. So, scalability issues occur in ODMRP.

Many modification techniques have been applied on ODMRP~\cite{Sung:1999} to improve the routing overhead. We can classify mesh based multicast routing protocol based on (1) modified routing mechanism and (2) on adding QoS parameter, for improvement in ODMRP protocol.
%%%%%%%%%%%%%%%%%%%%%%%%%%%%%%%%%%%%%%%%%%%%%%%%%%%%%%%%%%%%%%%%%%%%%%%%%%%%%%%%%%%%%%%%%%%%%%%%%%%%%%
\begin{figure*}
	\begin{center}
		\includegraphics[scale=0.3]{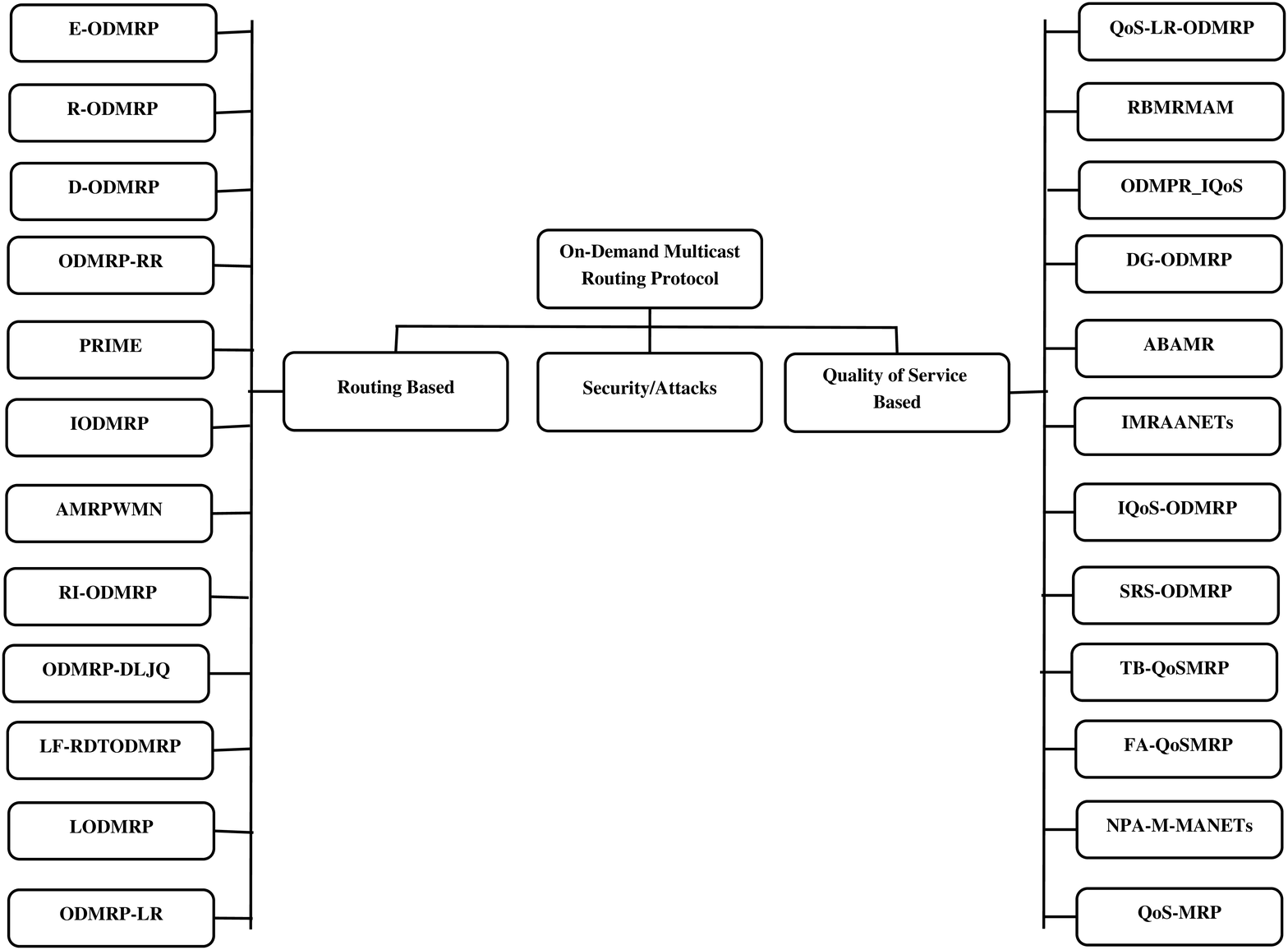} 
		\caption{\small{Enhancement in ODMRP on Routing and QoS}}
		\label{fig:taxonomy_odmrp}
	\end{center}
\end{figure*}

\subsection{Routing Based}
\label{Routing Based}
Routing mechanism in MRP is modified to make it more reliable and robust in terms of packet delivery ratio, end-to-end delay, control overhead and traffic load. Enhancements in the base ODMRP protocol are based on following routing modification approach:
\begin{enumerate}   
	\item Local Route Repair: This mechanism is used in order to avoid global broadcast of messages in case of route or link failure. Only broken link can demand for route and repair it by local recovery mechanism.
	\item Receiver Joining: In this mechanism, if any new incoming destination wants to join current route, it can request for a route from nearby forwarding node, multicast group or source by broadcasting request packet.
	\item Dynamic Timer Adaption: Motion adaptive refresh interval is utilizing link breakage report to source by receiver. Receiver can make adaptive interval according to their average link lifetime in route to make reliability.
	\item Periodic Hello: Periodic Hello packet is broadcast between nodes to extract neighbors' information or link quality.
	\item Route Discovery Suppression: It is used for limiting the number of simultaneous route discoveries as another discovery in process. Route Discovery Suppression~(RDS) helps us to reduce load on network.
	\item Conserving FG joining: In this mechanism, omit the joining of excessive number of Forwarding Group nodes in route to reduce overhead.
\end{enumerate}
These multiple mechanisms have been used to improve mesh based ODMRP protocol. In Figure~\ref{fig:taxonomy_odmrp}, we have listed routing modification protocols over ODMRP protocol.
%%%%%%%%%%%%%%%%%%%%%%%%%%%%%%%%%%%%%%%%%%%%%%%%%%%%%%%%%%%%%%%%%%%%%%%%%%%%%%%%%%%%%%%%%%%%%%%%%%%%%%
\subsubsection{ODMRP-MPR: On-Demand Multicast Routing Protocol with MultiPoint Relay in MANETs~\cite{Yao:2003}}
ODMRP-MPR inducts multipoint relay techniques to reduce the control overhead, obtain high stability and effectively solve the unidirectional link problem of wireless communication.
In network, each node N selects some neighbors on the basis of their distance from N and decides 2 hops as its multipoint relay~(MPR), only those neighbors will re-transmit the flooding packet broadcast by N.\\
ODMRP-MPR reduces flooding overhead generated by Join Query, re-transmission of Join Reply, avoids uni-directional link in forwarding path. But it increases additional overhead by sending periodic Hello messages. NS2 simulator has been used for simulation and compared the control overhead and PDR with varying number of senders and multicast group size with ODMRP protocol.
%%%%%%%%%%%%%%%%%%%%%%%%%%%%%%%%%%%%%%%%%%%%%%%%%%%%%%%%%%%%%%%%%%%%%%%%%%%%%%%%%%%%%%%%%%%%%%%%%%%%%%
\subsubsection{RODMRP: Resilient On Demand Multicast Routing Protocol~\cite{Pathirana:2007}}
The authors have offered more reliable forwarding path in case of node or route failure in mobility. The redundant packet forwarding improves PDR, while eliminating the possibility of flooding in networks. They create Non-Forwarding node~(NFG), that is not a member of Forwarding Group. It is further characterized into active non-forwarding node and passive non-forwarding node. Active non-forwarding nodes forward the data packet in network to improve the degradation in performance caused by node failure. It finds improvement in packet delivery ratio as compared with ODMRP on NS2 simulator.
%%%%%%%%%%%%%%%%%%%%%%%%%%%%%%%%%%%%%%%%%%%%%%%%%%%%%%%%%%%%%%%%%%%%%%%%%%%%%%%%%%%%%%%%%%%%%%%%%%%%%%
\subsubsection{EODMRP: Enhanced ODMRP with Motion Adaptive refresh~\cite{Oh:2008}}
Enhancement in ODMRP with refresh rate dynamically adapted to the environment and receiver joining. Receiver initiates join query (Receiver Join Query) in network to join a multicast group. If there exist a route to Multicast Receiver or Forwarding Group member, they should reply with Receiver Join Reply. Receiver increases TTL value and repeats process until the upper limit of TTL reaches. In worst case scenario, if no route is found, the receiver floods a refresh request packet.\\
Compares the variation in PDR and control overhead with increment in number of receivers with Qualnet simulator. Simulations show that E-ODMRP achieves higher PDR. Protocol has some limitations because it uses dummy packets and transmit to a sub-tree to prevent recovery explosion, which may result in extra overhead. It also increases extra routing overhead by sending Receiver Join Query~(RJQ) packet by receiver to join current route and needed additional processing power. Attacker can also waste their resources by sending numerous RJQ request.
%%%%%%%%%%%%%%%%%%%%%%%%%%%%%%%%%%%%%%%%%%%%%%%%%%%%%%%%%%%%%%%%%%%%%%%%%%%%%%%%%%%%%%%%%%%%%%%%%%%%%%
\subsubsection{AMRPWMN: Adaptive Multicast Routing Protocol for Wireless Mobile Adhoc Networks~\cite{Lai:2009}}
Improves control packet overhead by broadcasting Join-Query packets according to the current Packet Delivery Ratio. Due to excessive network overhead and collision, the protocol uses PDR to evaluate Join Query transmission. Sender in the network broadcasts join query packet according to probability variable, which is calculated by PDR. If PDR is high, AMRPWMN can transmit more Join Query, else there would be much collision in network. Simulation has been carried out on Glomosim 2.03 simulator and effect of variation of number of senders on different packet delivery ratio has been analysed.
%%%%%%%%%%%%%%%%%%%%%%%%%%%%%%%%%%%%%%%%%%%%%%%%%%%%%%%%%%%%%%%%%%%%%%%%%%%%%%%%%%%%%%%%%%%%%%%%%%%%%%
\subsubsection{LF-RDTODMRP: A Robust and Efficient On-demand Multicast Routing Protocol for Adhoc Network~\cite{Ghafouri:2009}}
This protocol cuts down the unnecessary redundant routes and their data transmission. They limit some nodes to flood Join-Request packets and forbid it to be a forwarding node. It adds a data structure (RDT table) for each forwarding node to make entry of multicast group, source address and time of entry. RDT table is used for reducing data transmission by using older route.
\par LF-RDTODMRP limits the flood requests of JQ packet. It adds a Load\_Table for storing number of times FG\_Flag has been set by multicast group. The protocol sets a threshold value for FG\_Flag. If the sum is greater than threshold, it drops the JQ packets. It sets the threshold value adaptive to the network for better output. This protocol has limitation of Hard to Selection of threshold for number of times FG\_Flag has been set. Simulations has been carried out on Glomosim Simulator and obtained results of RDTODMRP, LF-RDTODMRP and ODMRP are compared based on PDR, Delay and Overhead with varying traffic load and number of senders.
%%%%%%%%%%%%%%%%%%%%%%%%%%%%%%%%%%%%%%%%%%%%%%%%%%%%%%%%%%%%%%%%%%%%%%%%%%%%%%%%%%%%%%%%%%%%%
\subsubsection{ODMRP-LR: ODMRP with Link Failure Detection and Local Recovery mechanism~\cite{Tahan:2009}}
Addresses the problem of detecting link breakages and local recovery procedure in ODMRP. Tries to improve the disadvantage of E-ODMRP by reducing routing overhead. Two methods have been used for detecting link failure: first, by utilizing the knowledge of time intervals between data packets that are to be received. Second, by using hello packets or data packets in predefined interval. In Glomosim simulator, results have been analysed for PDR and control overhead on varying TTL value with mobility or non-mobility.
%%%%%%%%%%%%%%%%%%%%%%%%%%%%%%%%%%%%%%%%%%%%%%%%%%%%%%%%%%%%%%%%%%%%%%%%%%%%%%%%%%%%%%%%%%%%%%%%%%
\subsubsection{RBMRMAM: Relay-Based Multicast Routing in Multirate-aware MANETs~\cite{Qingshan:2009}}
Minimizes the total transmission time by extracting higher transmission rate of relay node. Proposes Heuristic Relay Node Selection Algorithm~(HRNSA) for choosing neighbor node that can use higher rate to cover more number of downstream node. It includes three modules: Information collection algorithm, Relay node selection algorithm and Relay notice algorithm. Throughput and delay are estimated with increasing number of nodes and speed of nodes and compared with that of ODMRP protocol on NS2 Simulator.
%%%%%%%%%%%%%%%%%%%%%%%%%%%%%%%%%%%%%%%%%%%%%%%%%%%%%%%%%%%%%%%%%%%%%%%%%%%%%%%%%%%%%%%%%%%%%%%%%
\subsubsection{ODMRP-RR: Multicast Routing Protocol for Reduction of Relay node in MANET~\cite{Nomata:2010}}
Reduces network overload by reducing number of relay nodes to enhance the performance of ODMRP. Tries to reduce the number of relay nodes that are used for constructing the route. Uses Round Robin scheduling for route construction for many sources. All the sources are not sending join query packets simultaneously. It uses round robin mechanism to differentiate between different sources and allots them distinct time slots. Computes average number of FG nodes for different number of source nodes on NS2 Simulator and Physical testbed.
%%%%%%%%%%%%%%%%%%%%%%%%%%%%%%%%%%%%%%%%%%%%%%%%%%%%%%%%%%%%%%%%%%%%%%%%%%%%%%%%%%%%%%%%%%%%%%%%%%%%%%
\subsubsection{ODMRP-DLJQ: Improving Performance of On-Demand Multicast Routing by Deleting Lost Join Query Packet~\cite{Abdollahi:2010}}
Improves the performance by restricting the domain of Join Query packet, which has been lost. Achieved by augmenting (increasing) the join query packet with minimum extra information which denotes the number of visited nodes from previous forwarding group nodes. If the current JQ visited many nodes and doesn't get any previous FG node, then discard it. Reduces overall overhead with increasing number of forwarding group and hop count over Glomosim Simulator.
%%%%%%%%%%%%%%%%%%%%%%%%%%%%%%%%%%%%%%%%%%%%%%%%%%%%%%%%%%%%%%%%%%%%%%%%%%%%%%%%%%%%%%%%%%%%%%%%%%%%%%
\subsubsection{PRIME: Interest-Driven Approach to integrated Uni-Multicast Routing~\cite{Garcia:2011}}
PRIME establishes meshes that are activated and deactivated by the presence or absence of interest in destinations and groups. PRIME establishes enclaves for flows of interest on-demand, and send proactively signals to update routing information within enclaves. Region of network with interest in the destination of flows receives timely updates as compared to other networks.
\par Meshes are activated using Mesh-activation Request (MR), which make receiver change their state from inactive to active state. The destination must start advertising its presence periodically using Mesh Announcements (MA). An enclave of multicast flow is a connected components that contains those node dissemination of information for flow. Analyzes group delivery ratio and delay with increasing number of group in Multicast and Unicast traffic for AODV, ODMRP and PUMA on Qualnet 3.9 Simulator.
%%%%%%%%%%%%%%%%%%%%%%%%%%%%%%%%%%%%%%%%%%%%%%%%%%%%%%%%%%%%%%%%%%%%%%%%%%%%%%%%%%%%%%%%%%%%%%%%%%%%%%
\subsubsection{RI-ODMRP: Receiver Initiated Mesh Based Multicasting for MANET using ACO~\cite{Deepalakshmi:2011}}
RI-ODMRP approach has been designed to find optimum paths between two communicating nodes. Initializes request by the node that wants to join the multicast group. An Ant Colony based mechanism is used for multicast routing protocol. Initialize/requesting node is named as core. Defines role of node by binary number 11, 01, 10, 00, where first byte is for forwarding group node and second one is for multicast group node. The process of route set up is performed in three steps: (1) Multicast Group Announcement, (2) Multicast Group Joining and (3) Join Reply. Evaluates average robustness and packet delivery ratio with mobility speed and compared with ODMRP on NS2 Simulator.
%%%%%%%%%%%%%%%%%%%%%%%%%%%%%%%%%%%%%%%%%%%%%%%%%%%%%%%%%%%%%%%%%%%%%%%%%%%%%%%%%%%%%%%%%%%%%%%%%%%%%%
\subsubsection{D-ODMRP: A Destination-driven ODMRP for MANETs~\cite{Yan:2012}}
\begin{figure}[h!]
	\begin{center}
		\includegraphics[scale=0.4]{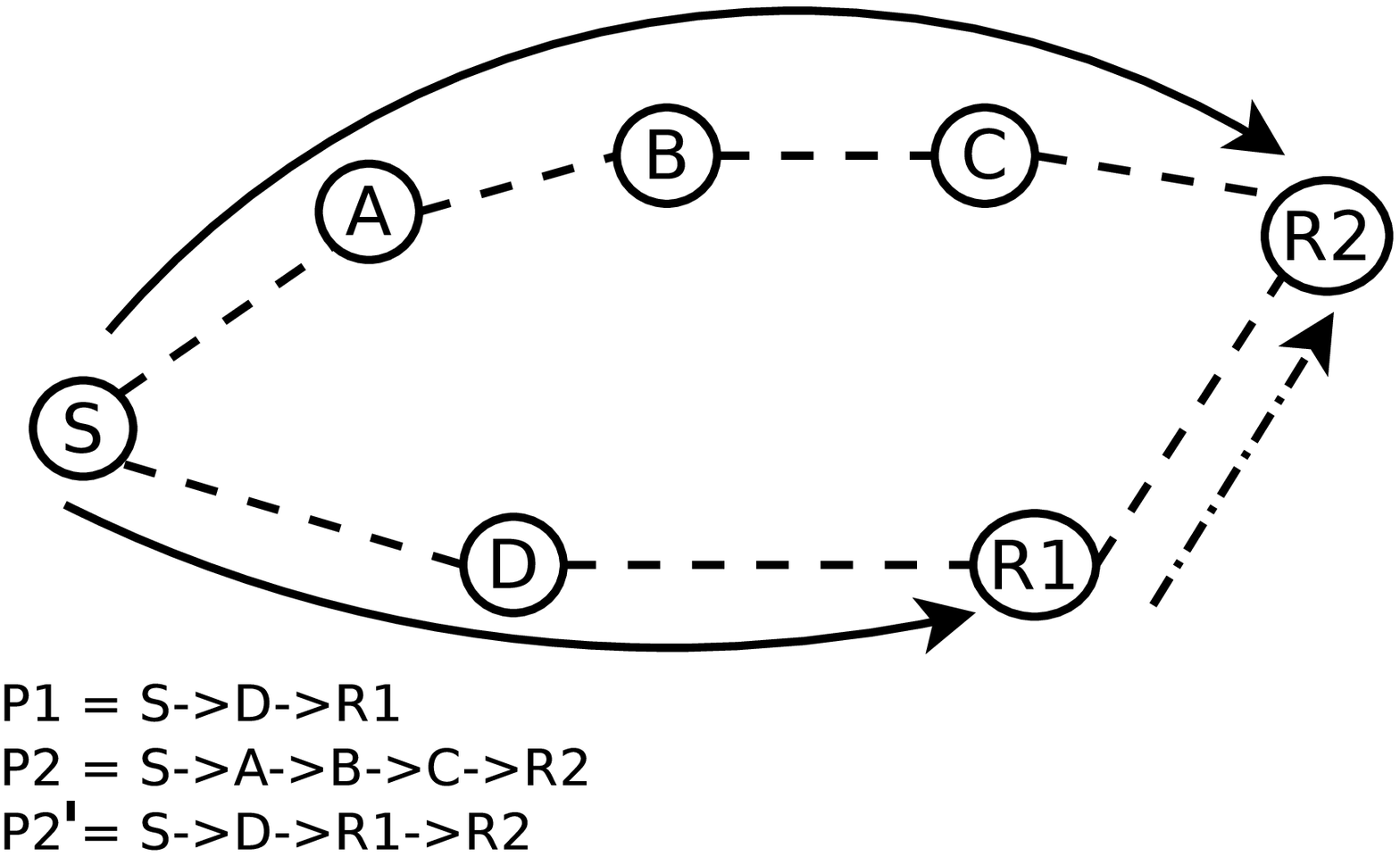} 
		\caption{\small{D-ODMRP Protocol}}
		\label{fig:dodmrp}
	\end{center}
\end{figure}

To improve the multicast forwarding efficiency in MANETs, D-ODMRP uses existing multicast destination node as forwarding node. In this protocol, the path from multicast source to multicast destination tends to use those paths passing through another multicast destination. In Figure~\ref{fig:dodmrp}, we have represented an example for one source and two receivers. In ODMRP protocol, paths P1 and P2 are selected, by default, for receivers R1 and R2 respectively. In D-ODMRP, R2 is nearer to R1 as compared to other receiver, so R2 can pick a route P2' via R1 as intermediate node and doesn't require any separate route.
\par If such multiple paths are available, the one leading to the least extra cost is preferred. It also takes deferring time to calculate delay for reaching packet to the destination. Simulated D-ODMRP for packet delivery ratio and control overhead and compared the results with that of ODMRP protocol over NS2 Simulator.
%%%%%%%%%%%%%%%%%%%%%%%%%%%%%%%%%%%%%%%%%%%%%%%%%%%%%%%%%%%%%%%%%%%%%%%%%%%%%%%%%%%%%%%%%%%%%%%%%%%%%%
\subsubsection{LODMRP: Level Based On-Demand Multicasting Routing Protocol for MANETs~\cite{Ghasemi:2013}}
Protocol tries to confine flooding of control packets within network by broadcasting only a part of these packet based on level-based approach. Each node decides to broadcast a Join Query packet based on its distance from the sender. The threshold for discarding join query is number of hops. Level represents the number of hops from sender to the node. Neighbor nodes transmit more packets as compared to far away nodes. Control overhead, efficiency and delay are analyzed with increasing number of sender and traffic load and compared results with ODMRP protocol over Glomosim Simulator.
%%%%%%%%%%%%%%%%%%%%%%%%%%%%%%%%%%%%%%%%%%%%%%%%%%%%%%%%%%%%%%%%%%%%%%%%%%%%%%%%%%%%%%%%%%%%%%%%%%%%%%
\subsection{QoS based}
\label{QoS Based}
\begin{figure}[h!]
	\begin{center}
		\includegraphics[scale=0.4]{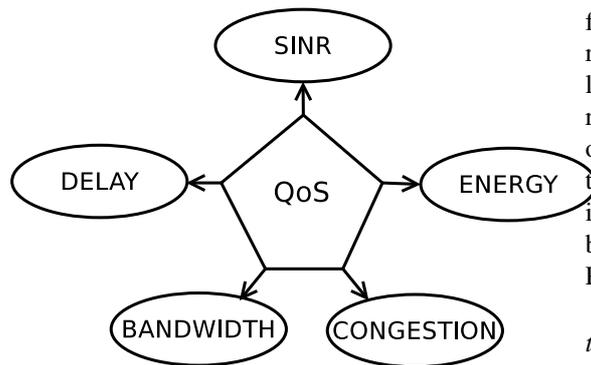} 
		\caption{\small{QoS Metrics}}
		\label{fig:QoS}
	\end{center}
\end{figure}

Quality of Service parameters are not used to discover path from source to destination, but to gratify the QoS requirements often given in terms of delay, congestion, bandwidth and power. In this section, we have discussed briefly the improvements that have been made in ODMRP protocol proposed by different authors to ensure QoS support and reliable in case of route or link breakage. In Figure~\ref{fig:taxonomy_odmrp}, we have listed QoS based modification protocols on ODMRP.
%%%%%%%%%%%%%%%%%%%%%%%%%%%%%%%%%%%%%%%%%%%%%%%%%%%%%%%%%%%%%
In Figure~\ref{fig:QoS}, we have represented QoS metrics that have been used by researchers for enhancement in ODMRP protocol, to make it more robust, reliable and reduce control overhead. These metrics have been used to make efficient route and less prone to link failure due to high stability links. Researchers always try to refine QoS metrics to improve delivery ratio without degrading network throughput.
%%%%%%%%%%%%%%%%%%%%%%%%%%%%%%%%%%%%%%%%%%%%%%%%%%%%%%%%%%%%%%%%%%%%%%%%%%%%%%%%%%%%%%%%%%%%%%%%%%%%%%
%%%%%%%%%%%%%%%%%%%%%%%%%%%%%%%%%%%%%%%%%
\subsubsection{QoS-LR-ODMRP: Quality of Service and Local Recovery for ODMRP Multicast routing in Adhoc Networks~\cite{EffatParvar:2007}}
Proposes a new technique for supporting QoS routing in ODMRP by making acceptable estimation of available and required bandwidth with local route discovery. Protocol ensures that every node in the route set up phase based on bandwidth calculations for available bandwidth. Consumed bandwidth of node channel is given by reserved bandwidth for flow on upstream and downstream neighbor of node. The protocol sets up route on the basis of available bandwidth of forwarding node. Protocol also proposes local route discovery on link breakage due to node mobility. Evaluates PDR and traffic admission ratio with increasing speed of node over Glomosim Simulator.
%%%%%%%%%%%%%%%%%%%%%%%%%%%%%%%%%%%%%%%%%%%%%%%%%%%%%%%%%%%%%%%%%%%%%%%%%%%%%%%%%%%%%%%%%%%%%%%%%%%%%%
\subsubsection{A cooperative framework for reliable multicast forwarding in MANETs~\cite{Araniti:2010}}
It offers higher reliability and connectivity among multicast members in comparison to other existing reactive protocols. Innovative framework based on the cooperation between MAC and routing protocol. It also adds some new features to ODMRP and IEEE 802.11 MAC layer for reliable forwarding. Added ODMRP with D3MP (Dynamic Mesh Based Multicast MAC Protocol) and RRAR (Round Robin Acknowledge and Re-transmit). Evaluates Signaling Overhead, PDR and Delay with variation in Multicast Group size on NS2 Simulator.
%%%%%%%%%%%%%%%%%%%%%%%%%%%%%%%%%%%%%%%%%%%%%%%%%%%%%%%%%%%%%%%%%%%%%%%%%%%%%%%%%%%%%%%%%%%%%%%%%%%%%%
\subsubsection{AAM-QoS: Agent Based Adaptive Multicast Routing with QoS guarantee in MANETs~\cite{Santhi:2010}}
Protocol guarantees QoS in terms of bandwidth, delay, jitter and packet loss with agent based adaptive algorithm. Set of static and mobile agent moves around the network and collects the routing information. Clustering of nodes and selection of QoS aware cluster head. Identifies intermediate node and discovers multiple paths to satisfy multiple constraints. Sets up a QoS aware path for the required multicast route. Evaluates packet delivery ratio and latency with mobility and group size on NS2 simulator.
%%%%%%%%%%%%%%%%%%%%%%%%%%%%%%%%%%%%%%%%%%%%%%%%%%%%%%%%%%%%%%%%%%%%%%%%%%%%%%%%%%%%%%%%%%%%%%%%%%%%%%
\subsubsection{IMRAANETs: An Improved Multicast Routing Algorithm in Adhoc Network~\cite{Zhang:2010}}
Analyzes the power variation of nodes to predict the topology change and link state. Calculates transmission power and rate of change of received power for any two intermediate nodes. Calculates the response time to inform source about unreliable link/node to prevent route failure. It reduces the route failure numbers and delivery delay without increasing extra overhead. Compares response time with failure time to trigger the routing warning function. Simulations have been carried out for both ODMRP and Extended protocol to evaluate PDR with varying mobility speed. %%%%%%%%%%%%%%%%%%%%%%%%%%%%%%%%%%%%%%%%%%%%%%%%%%%%%%%%%%%%%%%%%%%%%%%%%%%%%%%%%%%%%%%%%%%%%%%%%%%%%%
\subsubsection{TB-QoSODMRP: A Tree based QoS Multicast Routing Protocol for MANETs~\cite{Qabajeh:2011}}
\begin{figure}[h!]
	\begin{center}
		\includegraphics[scale=0.25]{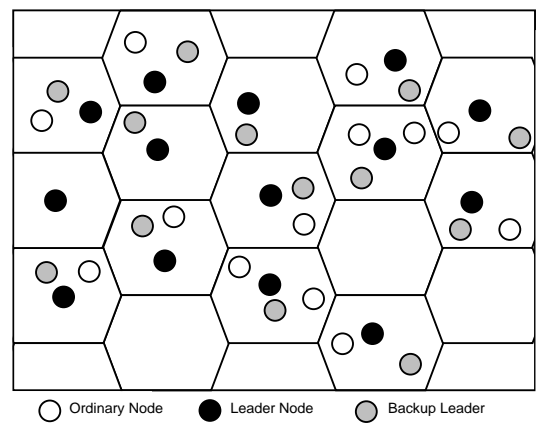} 
		\caption{\small{Hexagonal cells in a Scenario}}
		\label{fig:TB-QoS}
	\end{center}
\end{figure}

Proposes a model that searches for QoS guaranteed path for a single source to set of destinations. Physical area is partitioned into equal sized hexagonal cells as shown in Figure~\ref{fig:TB-QoS} and a leader and backup leader are elected to maintain updated information about the topology. Position based QoS Multicast Routing Protocol was proposed with GPS enabled on each node (device). The leaders are in the range of each node in the hexagonal cell. They find route on the basis of available bandwidth and delay to reach other intermediate nodes or destination. It is a type of group or cluster of nodes to transmit data effectively to each node by leader. Evaluates TBQMRP for CTRL packets transferred and packet loss ratio with mobility speed on Glomosim Simulator. But, there is a drawback in case of leader and backup leader. All communications that are performed, go under the leader node. So, the leader node has higher bandwidth and energy, else it would have gone through lack of resources.
%%%%%%%%%%%%%%%%%%%%%%%%%%%%%%%%%%%%%
%%%%%%%%%%%%%%%%%%%%%%%%%%%%%%%%%%%%%%%%%%%%%%%%%%%%%%%%%%%%%%%%%%%%%%%%%%%%%%%%%%%%%%%%%%%%%%%%%%%%%%
\subsubsection{ODMRP-IQoS: Providing Interference-aware Quality of Service Support for ODMRP~\cite{Yao:2009}}
Interference-aware QoS-ODMRP investigates bandwidth consumption under 2-hop interference model. Evaluates available bandwidth by employing a bit vector, named as Time Tag, to trace the transmission status within 2-hop neighbors. Finds clique (two or more nodes are in same transmission range) in network to avoid transmission interference. TTag is used for recording transmission status in most recent time from one hop neighbors. Nodes exchange TTAG with neighbors periodically for estimating bandwidth requirement. It analyzes delivery ratio and delay with increasing payload and shows improvement mainly with payload increment over Glomosim Simulator. Periodic transmission of TTag among neighbors is major con for the protocol performance.
%%%%%%%%%%%%%%%%%%%%%%%%%%%%%%%%%%%%%%%%%%%%%%%%%%%%%%%%%%%%%%%%%%%%%%%%%%%%%%%%%%%%%%%%%%%%%%%%%%%%%%
\subsubsection{IODMRP: Improvement of wireless multicast routing with Energy-efficiency based on ODMRP~\cite{Ying:2009}}
It takes partial nodes in forwarding group that relay packets and its choice is based on forwarder density and power state. Lesser the number of neighbor forwarding nodes, higher is the PDR. Some portion of forwarding group forwards the packet, where the portion is calculated on the basis of probability, p(0 $\le$ p $\le$ 1). Calculates power state (PS) of node by dividing current received power by initial power. Analyzes end-to-end delay and PDR of IODMRP over different number of receivers and maximum mobility speed over NS2 simulator.
%%%%%%%%%%%%%%%%%%%%%%%%%%%%%%%%%%%%%%%%%%%%%%%%%%%%%%%%%%%%%%%%%%%%%%%%%%%%%%%%%%%%%%%%%%%%%%%%%%%%%%
\subsubsection{DG-ODMRP: Delay-Guaranteed Multicast Routing in Multi-rate MANETs~\cite{Yu:2009}}
Estimates one hop delay and end-to-end delay based on varied transmission rates by monitoring the sensed busy to idle ratio of shared channel. Calculating both delays using IEEE 802.11 MAC is still challenging, because the radio channel is shared among neighbors. One hop delay is sum of deferring and transmission time. Protocol senses busy to idle ratio of shared channel for one hop delay. The end-to-end delay can be determined by summing up all one hop delays in route. This approach also considers link with maximum signal rate. Compares one hop and end-to-end delay with DGMR and AQOR in single and multi-rate environment on NS2 simulator.
%%%%%%%%%%%%%%%%%%%%%%%%%%%%%%%%%%%%%%%%%%%%%%%%%%%%%%%%%%%%%%%%%%%%%%%%%%%%%%%%%%%%%%%%%%%%%%%%%%%%%%
\subsubsection{LSMRM: Link stability multicast routing protocol in MANETs~\cite{biradar:2010}}
Authors select stable forwarding node that is based on link connectivity with high stability in mesh based multicast routing protocol. Stable route is selected by determining stable nodes which have high link quality in terms of estimated received power, bit error rate per packet and distance between communicating nodes. They have maintained link stability database at every node. The drawback is that they have not given any mathematical and analytical model to prove or validate their implementations and results. They have improved PDR, delay and routing overhead over changing multicast group size and transmission range and compared the obtained experimental results with ODMRP and EODMRP.
%%%%%%%%%%%%%%%%%%%%%%%%%%%%%%%%%%%%%%%%%%%%%%%%%%%%%%%%%%%%%%%%%%%%%%%%%%%%%%%%%%%%%%%%%%%%%%%%%%%%%%
\subsubsection{SRS-ODMRP: Stable Route Selection in ODMRP with Energy Based Strategy~\cite{Begdillo:2010}}
Stable Route Selection forwards data on the basis of node energy. To select stable route, route expiration time and residual energy have been considered. Stable Weight Based method is used for ODMRP protocol to improve reliability. Calculates Residue Energy (RES) and Route Expiration Time (RET), combines them to calculate shortest route. It appends position, direction, speed and mobility. It analyzes end-to-end delay and control overhead with variation in mobility speed and multicast group size on OPNET simulator.
%%%%%%%%%%%%%%%%%%%%%%%%%%%%%%%%%%%%%%%%%%%%%%%%%%%%%%%%%%%%%%%%%%%%%%%%%%%%%%%%%%%%%%%%%%%%%%%%%%%%%%
\subsubsection{IQoS-ODMRP: A novel routing protocol considering QoS Parameter  in MANETs~\cite{Jabbehdari:2010}}
Extends the ODMRP protocol to make it more suitable in disaster area network with group communication. Adds QoS parameters like bandwidth and delay in ODMRP. Takes consideration of mobility and analyses the effect of time interval of sending packet with change in mobility. Improves PDR and delay and compares the result with QoS-ODMRP on GlomoSim Simulator.
%%%%%%%%%%%%%%%%%%%%%%%%%%%%%%%%%%%%%%%%%%%%%%%%%%%%%%%%%%%%%%%%%%%%%%%%%%%%%%%%%%%%%%%%%%%%%%%%%%%%%%
\subsubsection{LLMR: A link stability-based multicast routing protocol for Wireless mobile adhoc network~\cite{Torkestani:2011}}
This protocol finds the longer route (stable multicast route) in high mobility scenario. Authors have used weighted multicast routing algorithm to generate stochastic Steiner tree within expected duration time~{EDT}. Then applied learning automata-based approach to solve the problem. They have done extensive simulation on NS2 and compared the result with LSMRM and EODMRP protocol to validate the results. They have calculated route life time and PDR with changing host speed and multicast group size. Improves PDR and delay. Compares the result with QoS-ODMRP over Glomosim Simulator.
%%%%%%%%%%%%%%%%%%%%%%%%%%%%%%%%%%%%%%%%%%%%%%%%%%%%%%%%%%%%%%%%%%%%%%%%%%%%%%%%%%%%%%%%%%%%%%%%%%%%%%
\subsubsection{FA-QoSMRP: Fuzzy Agent Based QoS Multicast Routing in MANETs~\cite{Budyal:2012}}
It provides the desired Quality of Service for user in group communication. A set of agents are used to operate in the following sequence. Creation of QoS Multicast mesh networks by using fuzzy inference system. A path to transmit the packet to receiver is selected from QoS Mesh. Mobile agents are employed to maintain QoS path. Analyzes PDR and control overhead and results reveal that FA-QoSMRP operates better than ODMRP on NS2 simulator.
%%%%%%%%%%%%%%%%%%%%%%%%%%%%%%%%%%%%%%%%%%%%%%%%%%%%%%%%%%%%%%%%%%%%%%%%%%%%%%%%%%%%%%%%%%%%%%%%%%%
\subsection{MMRNS: Neighbor Support reliable multipath multicast routing in MANETs~\cite{Biradar:2012}}
Authors have proposed a scheme for multipath multicast routing in MANETs using reliable neighbor selection. In this, a mesh is created from source to multicast destinations using maximum reliable pair factor of neighbors. In this algorithm, reliable pair factor depends upon energy and signal strength of node. Neighbor nodes are pruned by minimum threshold value and maintain route against node/link failure. Authors have analyzed their results with ODMRP and EODMRP protocol in terms of PDR and control/computation overhead in respect of number of nodes and groups with mobility considerations.
%%%%%%%%%%%%%%%%%%%%%%%%%%%%%%%%%%%%%
\begin{figure}[h!]
	\begin{center}
		\includegraphics[scale=0.35]{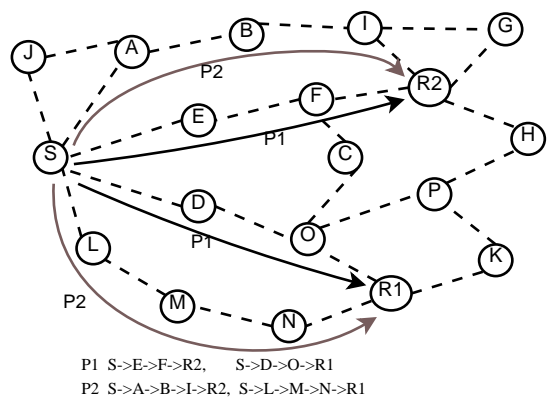} 
		\caption{\small{Multiple Multicast using Reliable Network Selection}}
		\label{fig:mmrns}
	\end{center}
\end{figure}
%%%%%%%%%%%%%%%%%%%
Figure~\ref{fig:mmrns} shows random network topology consisting of one source(S) and two destinations(R1 and R2). In this Figure, MMRNS picks up two reliable paths for each destination at any time t. Firstly, the source transmits data through higher priority level path P1 to destinations R1 and R2. In multicasting, load of the transmission has been increased on single node/link due to multiple transmissions. So, using different priority paths, load can be reduced by transmitting data through multiple paths like P2.	
%%%%%%%%%%%%%%%%%%%%%%%%%%%%%%%%%%%%%%%%%%%%%%%%%%%%%%%%%%%%%%%%%%%%%%%%%%%%%%%%%%%%%%%%%%%%%%%%%%%
\subsection{On-Demand multicast routing protocol with efficient route discovery~\cite{Mohammad:2012}}
Limited flooding in ODMRP reduces the packet overhead drastically, by sending JQ messages from only delay satisfaction nodes. Calculates one-hop delay for every node by summing up transmission delay, contention delay and queuing delay. Node selects minimum hop delay node to transmit the data and only floods Join Query messages. Analyzes delay, overhead and PDR with increasing number of multicast receivers over NS2 simulator and compares results with ODMRP and EODMRP protocol.
%%%%%%%%%%%%%%%%%%%%%%%%%%%%%%%%%%%%%%%%%%%%%%%%%%%%%%%%%%%%%%%%%%%%%%%%%%%%%%%%%%%%%%%%%%%%%%%%%%%%%%
\subsubsection{QoS-MRPM: QoS Based Multicast Routing Protocol in MANETs~\cite{Zheng:2012}}
QoS based MRP provides stable multicast paths with enough bandwidth. Entropy is treated as an important parameter to find stable path. Protocol uses bandwidth reservation mechanism to achieve QoS. It can be used to select stable path with longer lifetime. Compares the results of Average Delay and PDR with varying Velocity of Sending Packet for ODMRP over NS2 simulator.
%%%%%%%%%%%%%%%%%%%%%%%%%%%%%%%%%%%%%%%%%%%%%%%%%%%%%%%%%%%%%%%%%%%%%%%%%%%%%%%%%%%%%%%%%%%%%%%%%%%%%%
\subsubsection{Extending ODMRP for On-Site Deployments in Disaster Area Scenarios~\cite{Kirchhoff:2013}}
Extended the ODMRP protocol to make it more suitable in disaster area network with group communication. It is link quality based routing protocol that requires Hello packet transmissions. Firstly, they prioritize control messages and used Overhead Reduction Mechanism to provide better throughput in disaster areas. Evaluates packet loss ratio for GPS packet per node over NS2 simulator and tested it on Physical testbed.
%%%%%%%%%%%%%%%%%%%%%%%%%%%%%%%%%%%%%%%%%%%%%%%%%%%%%%%%%%%%%%%%%%%%%%%%%%%%%%%%%%%%%%%%%%%%%%%%%%%%%%
\subsubsection{NPA-MAM: New Power-aware multicast algorithm for mobile adhoc networks~\cite{Varaprasad:2013}}
Power aware multicast routing algorithm uses the residual battery life for multicasting from source to a group of destinations. Proposed protocol considers residual energy as a QoS metric while forwarding the data packets. The proposed model chooses a node with maximum remaining power among all the nodes. It extends the network lifetimes of the node and the network without degrading the network throughput. Compares the results with network life time, control bytes per data and PDR with varying group size on NS2 simulator.
%%%%%%%%%%%%%%%%%%%%%%%%%%%%%%%%%%%%%%%%%%%%%%%%%%%%%%%%%%%%%%%%%%%%%%%%%%%
% % % % % % % % % % % % % % % % % % % % % % % % % % % % % % % %
\subsection{LSMRP: Link Stability Multicast Routing Protocol in MANETs~\cite{Gaurav:2014}}
SINR is used for link stability estimation over ODMRP protocol. Algorithm determines reliable path in order to reduce link failures and re-routing overhead. It simply estimates a link which sustains for longer duration in the network. Link failure may occur due to mobility of nodes. Higher signal strength links have higher probability of existence. It increases route lifetimes without degrading the network throughput. Performed extensive simulations with increasing mobility and multicast group size on Exata simulator. Improved performance in terms of PDR and average End-to-End Delay.
\begin{figure}[h!]
	\begin{center}
		\includegraphics[scale=0.26]{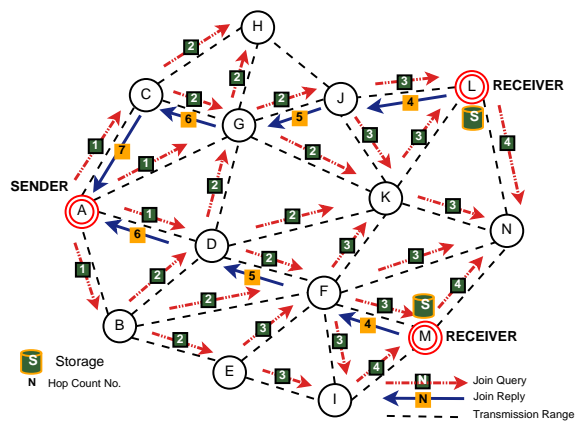} 
		\caption{\small{LSMRP Work Flow}}
		\label{fig:LSMRP}
	\end{center}
\end{figure}
\par We have estimated link stability of every link en-route from source to destination. To ensure that information from all available links is collected before selecting best possible link at a node, every packet is buffered for an `expiry time'. Unlike other methods which respond to first incoming packet, our method selects link with highest SINR as it is likely to be sustained for longer durations. In Figure~\ref{fig:LSMRP}, we have shown the route establishment with storage of packet at every receiving node. Steps in our proposal are as follows:\\
\begin{enumerate}
	\item Source generates a `Join Query' packet for multicast group address.
	\item Intermediate nodes receive `Join Query' from other nodes.
	\item Duplicate packets from same source to same multicast address and sequence number are discarded.
	\item Receiver node calculates SINR ratio from different senders, stores in a buffer with corresponding source address, sequence number and multicast address.
	\item Each node adds a timer that is initiated at receiving time of first packet. This packet is set to `Expiry Time' and only packets received in this duration are considered. 
	\item After `Expiry Time', a node calculates SINR of all links and selects one with maximum value. Also respective node is set as the previous hop address with maximum SINR ratio of incoming node.
	\item Intermediate node, then, broadcasts the packet for destination.
\end{enumerate}
\par According to the basic ODMRP protocol, minimum hop count is utilized to determine the paths between source and destination nodes. We made use of signal strength for path determination.
%%%%%%%%%%%%%%%%%%%%%%%%%%%%%%%%%%%%%%%%%%%%%%%%%%%%%%%%%%%%%%%%%%%%%%%%%%%%%%%%%%%%%%%%%%%%%%%%%%%%%%
%%%%%%%%%%%%%%%%%%%%%%%%%%%%%%%%%%%%%%%%%%%%%%%%%%%%%%%%%%%%%%%%%%%%%%%%%%%%%%%%%%%%%%%%%%%%%%%%%%%%%%
%\par The Table~\ref{tab:summary} presents the summary of mesh based multicast routing protocols on the basis of their extensions such as modification parameters, features, techniques and improvement parameters.
\par The table 1 represents the summary of mesh based multicast routing protocols. In the table, protocols are classified in the given components such as modification components, features, techniques and improvement parameters.
\begin{landscape}
	\begin{table*}[!htbp]
		%       \label{tab:summary}
		%    \vspace*{2.5cm}
		%    \hspace*{-30cm}
		% Landscape page
		
		%	\begin{table}[htbp!]
		\centering % Center table
		%\begin{table*}
		%\begin{sidewaystable}
		\scalebox{1}{
			%	\rotatebox{90}{
			\hspace*{-8cm}
			
			\begin{tabular}{|l|l|l|l|l|l|l|l|l|l|l|l|l|l|l|l|l|l|l|l|l|l|l|l|l|l|} \hline
				%               \label{tab:summary}
				%        \hspace*{-10cm}        
				\textbf{\large{Protocol Name}} &
				\textbf{\begin{turn}{-90}ODMRPMPR\end{turn}} & \textbf{\begin{turn}{-90}R-ODMRP\end{turn}} & \textbf{\begin{turn}{-90}EODMRP\end{turn}} &  \textbf{\begin{turn}{-90}AMRPWMN\end{turn}} & \textbf{\begin{turn}{-90}LF-RDTODMRP\end{turn}} & \textbf{\begin{turn}{-90}ODMRP-LR\end{turn}} & \textbf{\begin{turn}{-90}RB-MRMAM\end{turn}} &	  \textbf{\begin{turn}{-90}ODMRP-RR\end{turn}} & \textbf{\begin{turn}{-90}ODMRP-DLJQ\end{turn}} & \textbf{\begin{turn}{-90}PRIME\end{turn}} &   \textbf{\begin{turn}{-90}RI-ODMRP\end{turn}} &  \textbf{\begin{turn}{-90}DODMRP\end{turn}} &    \textbf{\begin{turn}{-90}LODMRP\end{turn}} & \textbf{\begin{turn}{-90}QoS-LR-ODMRP\end{turn}} &	  \textbf{\begin{turn}{-90}ODMRP-IQoS\end{turn}} & \textbf{\begin{turn}{-90}DG-ODMRP\end{turn}} & \textbf{\begin{turn}{-90}AAM-QoS\end{turn}} &	  \textbf{\begin{turn}{-90}IMRAANETs\end{turn}} & \textbf{\begin{turn}{-90}IQoS-ODMRP\end{turn}} & \textbf{\begin{turn}{-90}SRS-ODMRP\end{turn}} &   \textbf{\begin{turn}{-90}TB-QoSMRP\end{turn}} &  \textbf{\begin{turn}{-90}MMRNS\end{turn}} &
				\textbf{\begin{turn}{-90}FA-QoSMRP\end{turn}} &    \textbf{\begin{turn}{-90}NPA-MAM\end{turn}} &
				\textbf{\begin{turn}{-90}QoS-MRPM\end{turn}}\\ \hline \hline
				%	\textbf{Protocol} &	 \textbf{\rotatebox[origin=c]{90}{ODMRP-MPR}} & \textbf{RODMRP} &    \textbf{\rotatebox[origin=c]{90}{EODMRP}} &   \textbf{\rotatebox[origin=c]{90}{AMRPWMN}} & \textbf{\rotatebox[origin=c]{90}{LF-FDTODMRP}} & \textbf{\rotatebox[origin=c]{90}{ODMRP-LR}} & \textbf{\rotatebox[origin=c]{90}{RB-MRMAM}} & \textbf{\rotatebox[origin=c]{90}{ODMRP-RR}} & \textbf{\rotatebox[origin=c]{90}{ODMRP-DLJQ}} & \textbf{\rotatebox[origin=c]{90}{PRIME}} &   \textbf{\rotatebox[origin=c]{90}{RI-ODMRP}} &  \textbf{\rotatebox[origin=c]{90}{DODMRP}} &    \textbf{\rotatebox[origin=c]{90}{LODMRP}} \\ \hline \hline
				%\textbf{}&\textbf{Profiling}&\textbf{Parameter}&\textbf{Modeling}&\textbf{Employed}&\textbf{}&\textbf{}\\ \hline
				%            Shun \textit{et al}~(2011)~\cite{Shun:2011} & Live & System--calls & Graph & -- & Detecting module--based malware using MapReduce  & Specific to only dll--based malware\\ 
				%             &  &  & based &  & Focus on persistent behavior of malware on servers  & The  \\ \hline
				\textbf{Year of Publication} & 2003 & 07 & 08 & 09 & 09 & 09 & 09 & 10 & 10 & 12 & 11 & 12 & 13 & 08 & 09 & 09 & 10 & 10 & 10 & 10 & 11 & 12 & 12 & 13 & 12 \\ \hline
				%	\textbf{Reference No:} &~\cite{Yao:2003}& ~\cite{Pathirana:2007} & ~\cite{Oh:2008} & ~\cite{Lai:2009} & ~\cite{Ghafouri:2009} & ~\cite{Tahan:2009} & ~\cite{Qingshan:2009} & ~\cite{Nomata:2010} & ~\cite{Abdollahi:2010} & ~\cite{Garcia:2011} & ~\cite{Deepalakshmi:2011} & ~\cite{Yan:2012} & ~\cite{Ghasemi:2013} & ~\cite{EffatParvar:2007} & ~\cite{Yao:2009} & ~\cite{Yu:2009} & ~\cite{Santhi:2010} & ~\cite{Zhang:2010} & ~\cite{Jabbehdari:2010} & ~\cite{Begdillo:2010} & ~\cite{Qabajeh:2011} &~\cite{Biradar:2012} & ~\cite{Budyal:2012} & ~\cite{Varaprasad:2013} & ~\cite{Zheng:2012} \\ \hline \hline
				%& & & & & & & & & & & & & & & & & & & & & & & & & 
				\textbf{Modification Components:} \\ \hline
				\textbf{Forwarding Group} &\cmark& \cmark & \cmark & \textendash &\textendash &\textendash & \cmark & \cmark & \cmark &\textendash &\textendash & \textendash & \cmark &\textendash &\textendash &\textendash &\textendash &\textendash &\textendash &\textendash &\textendash &\textendash &\textendash &\textendash & \cmark \\ \hline
				\textbf{Routing Selection} &\cmark& & & \cmark &\textendash &\textendash &\textendash &\textendash & \cmark & & \cmark & \textendash& \cmark & \cmark & \cmark & \cmark & \cmark & \cmark & \cmark & \cmark & \cmark & \cmark&\cmark & \cmark & \cmark \\ \hline \hline
				\textbf{Features:} %& & & & & & & & & & & & & & & & & & & & & & & & & 
				\\ \hline
				\textbf{Local Route} &\textendash&\textendash & \cmark &\textendash &\textendash & \cmark &\textendash &\textendash &\textendash &\textendash &\textendash &\textendash &\textendash & \cmark &\textendash &\textendash &\textendash &\textendash & \textendash&\textendash &\textendash &\textendash &\textendash &\textendash & \textendash \\ 
				\textbf{Repair} & & & & & & & & & && & & & & & & & & && && &&  \\ \hline
				\textbf{Receiver Joining} &\textendash& \textendash& \cmark & \textendash& \textendash& \cmark & \textendash& \textendash& \textendash& \cmark & \textendash& \textendash& \textendash& \textendash& \textendash& \textendash& \textendash& \textendash& \textendash& \textendash& \textendash&\textendash & \textendash& \textendash& \textendash \\ \hline	
				\textbf{Dynamic Timer} &\textendash&\textendash & \cmark &\textendash &\textendash &\textendash &\textendash &\textendash &\textendash &\textendash &\textendash &\textendash & \cmark &\textendash &\textendash &\textendash &\textendash &\textendash & \cmark &\textendash &\textendash &\textendash &\textendash &\textendash &\textendash \\ 
				\textbf{Adaption} & & & & & & & & & && && & & & & & & & & && &&  \\ \hline
				\textbf{Mobility Handling} &\textendash&\textendash &\textendash &\textendash & \cmark &\textendash &\textendash &\textendash &\textendash & \cmark &\textendash &\textendash &\textendash &\textendash &\textendash &\textendash &\textendash &\textendash & \cmark & \cmark &\textendash &\cmark&\textendash &\textendash &\textendash  \\ \hline
				\textbf{Link Asymmetry} &\cmark&\textendash &\textendash &\textendash &\textendash &\textendash &\textendash &\textendash &\textendash &\textendash &\textendash &\textendash &\textendash &\textendash &\textendash &\textendash &\textendash &\textendash&\textendash &\textendash &\textendash &\textendash &\textendash &\textendash &\textendash  \\ \hline
				\textbf{Energy} &\textendash&\textendash &\textendash &\textendash &\textendash &\textendash &\textendash &\textendash &\textendash &\textendash &\textendash &\textendash &\textendash &\textendash &\textendash &\textendash &\textendash & \cmark & & \cmark & \cmark &\cmark& \cmark & \cmark &\textendash  \\ \hline
				\textbf{Delay} &\textendash&\textendash &\textendash &\textendash &\textendash &\textendash &\textendash &\textendash &\textendash &\textendash & \textendash&\textendash &\textendash &\textendash &\textendash & \cmark & \cmark & & \cmark & \cmark &\textendash &\textendash & \cmark &\textendash &\textendash  \\ \hline
				\textbf{Signal Strength} &\textendash& \textendash& \textendash& \textendash& \textendash& \textendash& \cmark & \textendash& \textendash& \textendash& \textendash& \textendash& \textendash&\textendash &\textendash & \cmark &\textendash &\textendash &\textendash &\textendash & \textendash& \cmark& \textendash& \textendash&\textendash  \\ \hline
				\textbf{Bandwidth Reservation} &\textendash&\textendash &\textendash &\textendash &\textendash &\textendash & \cmark &\textendash &\textendash &\textendash &\textendash &\textendash &\textendash & \cmark & \cmark & \textendash& \cmark & \textendash& \cmark &\textendash &\textendash &\textendash &\textendash &\textendash &\textendash  \\ \hline
				\textbf{Congestion/Interference} &\textendash& \textendash&\textendash &\textendash &\textendash &\textendash &\textendash &\textendash &\textendash &\textendash & \textendash&\textendash &\textendash & \cmark & \cmark & \textendash& \textendash& \textendash& \textendash& \textendash& \textendash&\textendash & \textendash& \textendash& \textendash\\ \hline
				\textbf{Application feedback} &\textendash&\textendash &\textendash &\textendash &\textendash &\textendash &\textendash &\textendash &\textendash &\textendash &\textendash &\textendash &\textendash &\textendash &\textendash &\textendash &\textendash &\textendash & \cmark &\textendash &\textendash &\textendash &\textendash &\textendash &\textendash \\ \hline \hline
				\textbf{Techniques:} %& & & & & & & & & & & & & & & & & & & & & & & & & 
				\\ \hline
				\textbf{Periodic Hello} &\cmark&\textendash &\textendash & \cmark &\textendash & \cmark & \textendash&\textendash & \textendash& \textendash&\textendash &\textendash &\textendash & \cmark & \cmark & \textendash& \textendash& \textendash& \cmark & \textendash& \textendash&\textendash &\textendash &\textendash &\textendash \\ \hline
				\textbf{Link Quality} &\textendash&\textendash & \cmark & \textendash& \textendash& \cmark &\textendash &\textendash &\textendash &\textendash & \textendash& \textendash& \textendash& \textendash& \textendash& \textendash& \textendash& \textendash& \textendash& \textendash& \textendash& \cmark& \textendash& \textendash& \textendash \\ \hline
				\textbf{Passive Ack} &\cmark& \textendash& \cmark &\textendash &\textendash & \textendash&\textendash &\textendash &\textendash &\textendash &\textendash &\textendash &\textendash &\textendash &\textendash &\textendash &\textendash &\textendash &\textendash &\textendash & \textendash&\textendash & \textendash& \textendash& \textendash \\ \hline
				\textbf{Join Query} &\textendash& \textendash&\textendash & \cmark &\textendash &\textendash & \textendash& \textendash&\textendash &\textendash &\textendash & \cmark & \textendash& \textendash&\textendash &\textendash &\textendash &\textendash &\textendash &\textendash &\textendash &\textendash &\textendash &\textendash &\cmark \\
				\textbf{Delaying} && & & & & & & & && & && & & & & & & & && &&  \\ \hline
				\textbf{Route Discovery} &\textendash&\textendash &\textendash & \cmark & \cmark &\textendash &\textendash &\textendash & \cmark &\textendash &\textendash &\textendash & \cmark &\textendash &\textendash &\textendash &\textendash & &\textendash &\textendash &\textendash &\textendash &\textendash &\textendash &\cmark \\
				\textbf{Suppression} & & & && & & & & & && & & & & & & & & & && &&  \\ \hline
				\textbf{Conserving FG} &\textendash&\textendash &\textendash &\textendash & \cmark &\textendash &\textendash & \cmark &\textendash &\textendash &\textendash & \cmark & \textendash& \textendash& \textendash& \textendash& \textendash& \textendash& \textendash& \textendash& \textendash& \textendash &\textendash& \textendash& \textendash \\ 
				\textbf{Joining} & & & & & & & & & && && & & & & & & & & && &&  \\ \hline \hline
				\textbf{Improvement Parameter:} %& & & & & & & & & & & & & & & & & & & & & & & & & 
				\\ \hline
				\textbf{PDR} &\cmark& \cmark & \cmark & \cmark & \cmark &\textendash & \cmark & \cmark &\textendash & \cmark & \cmark &\textendash & \cmark & \cmark & \cmark &\textendash & \cmark & \cmark & \cmark & \textendash& \textendash&\cmark& \cmark &\textendash & \cmark \\ \hline
				\textbf{Delay} &\textendash& \textendash& \textendash& \textendash& \cmark & \textendash& \cmark &\textendash & \cmark & \cmark & \textendash& \textendash& \cmark &\textendash & \cmark & \cmark & \cmark & \textendash& \cmark & \cmark &\textendash &\textendash &\textendash & \cmark & \cmark \\ \hline
				\textbf{Overhead} &\textendash& \textendash& \textendash& \textendash& \cmark & \cmark & \textendash& \textendash& \cmark &\textendash &\textendash & \cmark & \cmark & \textendash&\textendash &\textendash &\textendash &\textendash &\textendash & \cmark & \cmark &\cmark& \textendash& \textendash& \textendash \\ \hline \hline
				
				%\& & %& & & & & & & & & & & & & & & & & & & & & & & & & 
				%\\ \hline
				\textbf{Category} &A& A & A/B & A/B & A/B & B & C & A & B & A & A & B & B & A/C & C & C & C & C & C & C & C & C & C & C & C \\ \hline \hline
				
				%& & %& & & & & & & & & & & & & & & & & & & & & & & & & 
				%\\ \hline
				\textbf{Simulator} &1& 1 & 3 & 2 & 2 & 2 & 1 & 1 & 2 & 3 & 1 & 1 & 2 & 2 & 2 & 1 & 1 & 1 & 2 & 4 & 2 &1 & 1 & 1 & 1  \\ \hline
				
				%	\textbf{Forwarding Group} & \checkmark & \cmark & \textendash & 2009 & 2009 & 2009 & 2009 & 2010 & 2010 & 2012 & 2011 & 2012 & 2013 \\ \hline
				
				%	Beaucamps \textit{et al} & PIN & Library & Feature & Formal & Malware detection based on suspicious keylogging behavior & Detect a shallow~(local) form of malware behavior \\ 
				%(2010)~\cite{Beaucamps:2010}&  & Calls & Statistics & Method & Trace execution and behavior patterns are modeled using FSA & Behavior patterns are interleaved during execution trace modeling \\ \hline
				%	Canali \textit{et al} & Anubis & System & $n$--gram & Formal & llllll & jjjj \\ 
				%(2012)~\cite{Canali:2012}&  & Calls & based & Method & mmmmm & kkkkk \\ \hline
			\end{tabular}
		}
		%       \label{tab:summary}
		\caption{\large{Summary of Mesh Based Multicast Routing Protocol(\cmark \@ Supported and \textendash \@ Not\_Supported)}}% Add 'table' caption
		\label{tab:summary}
		%\end{table}
		%\end{sidewaystable}
	\end{table*}
\end{landscape}
\section{Introduction to Link Stability in Multicast Routing Protocol}
\label{linkstability}
Quality of Service is calculated on the basis of link stability. Fluctuating link stability in wireless networks has a fundamental impact on network performance. Link stability is subject to less robust link/node failure, node mobility and low reliability. Researchers try to make efficient routing protocols that are able to deal with link unreliability. Route persists for longer duration because stable links are en-routed. As a result, reduced computations due to less re-routings lead to reduced overhead. Here, we are giving a brief overview of requirement of link stability and its metrics.

%It reduces routing overhead by less computation in reconfiguration, due to long time existing route.

\subsection{Requirement of Link Stability}
In this section, we are going to discuss the requirement or benefits of adding the concept of link stability in the route construction. We have listed some of them below:
\begin{itemize}
	\item Energy Efficient
	%\par We can save Energy from reducing number of communication and computation on nodes by reducing number of reformation. Nodes energy is waste due to request, reply packets for demanding of route again and again in multiple link failure.
	\par Energy can be conserved by reducing number of reformations, i.e. selecting a stable route. Node energy is wasted due to broadcast of request and reply packets for repeated demand of route, in case of multiple links failure.
	\item Accuracy
	%\par We are finding longer/stable path from single discovery, so we have to ensure that link is not break from any reason.
	\par It is essential to estimate a longer/stable path from single discovery in order to ensure that link would not break for any reason.
	\item Reactivity
	\par Due to high mobility in the network, our protocol should be well adaptive to every small change in the network.
	\item Stability
	%\par We have to select stable route is terms of longer time. Nodes which have minimum distance, maximum power are more stable compare to other.\\
	\par Stable route in terms of prolonged persistence should be selected. Nodes at minimum distance and with maximum residual power are regarded as more stable than others. \\
\end{itemize}
There is a flaw between stability and reactivity that they can't be performed together because both are opposite to each other. So, we have to make reactive protocols that satisfy both conditions; stabilization and reactivity.
%%%%%%%%%%%%%%%%%%%%%%%%%%%%%%%%%%%%%%%%%%%%%%%%%%%%%%%%%%%%%%%%%%%%%%%%%%%%%%%%%%%%%%%%%%%%%%%%%%%%%%
\subsection{Link Stability Metrics}
Link stability metrics help us to find QoS aware links. Metrics that affect the link stability are as follows:
\begin{itemize}
	\item Signal Strength or SINR or BER
	\item Transmission Delay or ETX
	\item Residual Energy or Power
	\item Bandwidth Reservation
	\item Congestion or Interference
	%		\item ETX or Transmission Count 
\end{itemize}
%%%%%%%%%%%%%%%%%%%%%%%%%%%%%%%%%%%%%%%%%%%%%%%%%%%%%%%%%%%%%%%%%%%%%%%%%%%%%%%%%%%%%%%%%%%%%%%%%%%%%%
\subsubsection{Signal Strength/SINR/BER}
SINR is the power of certain signal of interest divided by sum of interference power from other signal and background noise. SINR is estimated from signal strengths between nodes for transmitting data.
%		$\textrm{SINR } = \frac{P}{I + N}$\@
\begin{equation}
%\begin{center}
\textrm{SINR} = \frac{Rec\_signal\_Power}{other\_Interference + background\_noise}
%\end{center}
\end{equation}
Bit Error Rate~(BER) can be decided from number of bits dropped out of total number of bits sent to the destination node. SINR is used to find stable link that has more lifetime than other nodes. We can also take continuous SINR values to predict the direction or mobility of node.
%%%%%%%%%%%%%%%%%%%%%%%%%%%%%%%%%%%%%%%%%%%%%%%%%%%%%%%%%%%%%%%%%%%%%%%%%%%%%%%%%%%%%%%%%%%%%%%%%%%%%%
\subsubsection{Delay/ETX}
Total Delay can be calculated by summing up the Transmission delay~(packet transfer time between nodes), Queuing delay~(packet has to wait in queue for getting sequenced) and Processing delay~(after queuing time for transferring the packet or wait for channel).
\begin{equation} %\centering
%\begin{center}
Delay = TD + QD + CD
%\end{center}
\end{equation}
where, TD is Transmission\_Delay, QD is Queuing\_Delay and CD is Contention\_Delay.
Transmission delay is the time taken to transmit a packet from one hop to next hop. In QoS metric, link with lower transmission delay will be included in the route for establishing stable path. Queuing delay and processing delay are approximately equal to all or otherwise it depends upon the type of application being executed.
%%%%%%%%%%%%%%%%%%%%%%%%%%%%%%%%%%%%%%%%%%%%%%%%%%%%%%%%%%%%%%%%%%%%%%%%%%%%%%%%%%%%%%%%%%%%%%%%%%%%%%
\subsubsection{Residual Energy/Power}
Energy has always been a major concern in MANETs. We are calculating node residual energy and energy consumption for $n$ transmissions. Most of the times, we assume that all nodes have same transmission power. Node which has more residual energy and less power consumption in data transmission is selected as intermediate node for transmission on route.
\begin{equation}
\textrm{Power Ratio} = \frac{Remaining Power}{Total Capacity}
\end{equation}
Power ratio tells us about node's remaining power. We can calculate power ratio by dividing remaining power by total capacity and conclude that it lies between two ranges; Low range and High range. Values above threshold comes in High power range.
%%%%%%%%%%%%%%%%%%%%%%%%%%%%%%%%%%%%%%%%%%%%%%%%%%%%%%%%%%%%%%%%%%%%%%%%%%%%%%%%%%%%%%%%%%%%%%%%%%%%%%
\subsubsection{Bandwidth reservation}
Network bandwidth reservation is used to identify the capacity of participating node and the corresponding link. We require minimum bandwidth to communicate data for QoS aware routing protocols. Available bandwidth of a node can be estimated through the idle time of channel and transmission range. We can calculate idle time of channel by monitoring the node, whether it is busy or not in receiving or sending any packet. Bandwidth reservation is compulsory in multimedia applications for improving delay and jitter in data streaming. For better analysis, we have to estimate network bandwidth precisely and total bandwidth consumption in transmission from requesting applications.
%%%%%%%%%%%%%%%%%%%%%%%%%%%%%%%%%%%%%%%%%%%%%%%%%%%%%%%%%%%%%%%%%%%%%%%%%%%%%%%%%%%%%%%%%%%%%%%%%%%%%%
\subsubsection{Congestion}
Congestion can be estimated in terms of interference and load on a node. We can calculate path encounter for every node to estimate congestion on a link or a node. Path encounter can be detected when a node comes in the transmission range of participating node. Path encounter value is the number of nodes in the transmission range~\cite{Son:2014} or number of nodes that are affecting~(consuming) network bandwidth of participating node. We can select node with minimum path encounter value at the time of route selection. Occurrence of congestion depends on various parameters. It can be related to bandwidth consumption and conflict due to simultaneous transmissions.

%Path encounter is the number of nodes in the transmission range~\cite{Son:2014} or nodes that affecting(consuming) network bandwidth of participating node. We can select node which have minimum path encounter value at time of route selection. Occurrence of congestion is depend on various parameter. It can be related to bandwidth consumption, conflict due to simultaneous transmission.
%%%%%%%%%%%%%%%%%%%%%%%%%%%%%%%%%%%%%%%%%%%%%%%%%%%%%%%%%%%%%%%%%%%%%%%%%%%%%%%%%%%%%%%%%%%%%%%%%%%%%%%%%%%%%%%%%%%%%%%%%%%%%%%%
\begin{figure}[h!]
	\begin{center}
		\includegraphics[scale=0.4]{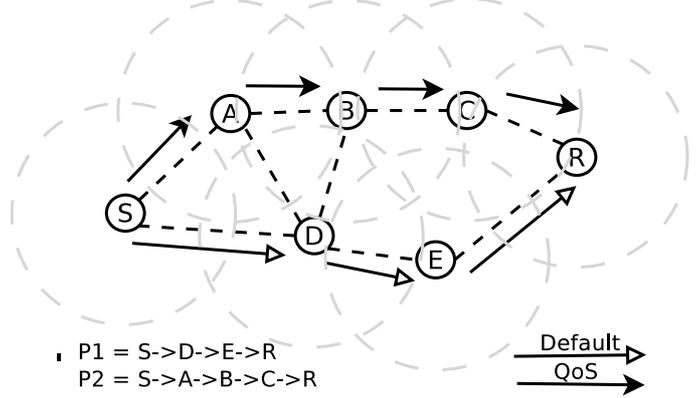} 
		\caption{\small{QoS\_Route Construction}}
		\label{fig:QoS_Route}
	\end{center}`
\end{figure}
\par In Figure~\ref{fig:QoS_Route}, we have represented a QoS aware route and a default route construction. We have two routes P1 and P2 as Default and QoS aware routes respectively. We can use QoS metrics to calculate stable link for transmitting data. In basic ODMRP protocol, receiver R can be reached via nodes D and E as forwarding group member, due to less hop count and fastest response from P1 route. But, we can not say that it is stable route or not. We can use QoS metric to calculate link availability time for finding stable and robust link for route construction as shown in P2. Although, route P2 has increased number of hops, yet entire route would remain stable till route expiration time.

\section{Design Issues and Challenges in Multicast Routing Protocol}
\label{Issue and challenges}
Although researchers have designed numerous multicast routing protocols and techniques, yet there remains a number of open issues and challenges.
% The suitability of any such approach depends upon three components:
%\begin{itemize}
%	\item	\textit{Completeness:}  Current state--of--the--arts in MRP blindly believe some assumption that may be dangerous for proposed mechanism. We have to developed the protocol that have high robustness and reliable communication.
%	\item	\textit{Correctness:} It should be correct in each scenario and 
%	
%	\item	\textit{Complexity:} Protocol should be not so much complex or simple
%	
%\end{itemize}
All available protocols provide mechanism to improve the performance of ODMRP protocol. However, each of these approaches suffer from certain issues which decide the preciseness of the approach. Some of the issues~\cite{Mohapatra:2005} are listed as follows:
\begin{itemize}
	\item[(i)] \textit{\textbf{Energy Efficiency:}} Mobile nodes typically run on limited energy resources, so it is required to design energy preserving protocols for group communication. Low throughput and high interferences over wireless channels is due to high energy consumption in MANETs.
	\par We can reduce energy consumption by decreasing number of nodes that are included in forwarding group and selecting the nodes that have highest energy to transmit the datagram, to make equal consumption by every node. We can also preserve energy through broadcast communication.
	\par Many protocols have been proposed for energy conservation~\cite{Ying:2009}. If we select a high-energy link in advance, then energy would not be consumed due to retransmission of packet at MAC layer. Energy issues increase with high mobility, high contention because we are using shared channel based MAC protocol.	
	\item[(ii)] \textit{\textbf{Robust and Reliable Multicasting:}} Due to arbitrary movement of mobile nodes, link failures are usual in MANETs. MRP should be resistive to mobility and gain high PDR. Reliable multicasting ensures that data from source node should reach every destination with ditto set of messages.
	\par In MANETs, the reliability of multicast frames cannot be guaranteed because it depends upon mobility, multicast group size and traffic load. In multicast routing protocols, there are no RTS/CTS control frames to enquire about the availability of channel. Moreover, there is no provision of acknowledgement to achieve reliable communication. Unreliability in the network increases due to transmission of real time multimedia traffic. A node is unstable or unreliable due to its high and unpredictable mobility.	
	\item[(iii)] \textit{\textbf{Efficiency and Control Overhead}} Efficiency can be defined as ratio of total number of packets received to total number of packets transmitted at receiver. Total number of control packets transmitted in the entire network to maintain routes to multicast group signify control overhead. Bandwidth consumption at a node is higher due to control packets like transmission of hello packets, route request packets etc.
	\item[(iv)] \textit{\textbf{QoS Aware Multicasting:}} QoS is achieved by set of service parameters during data streaming over multicast group from a source. QoS attributes like delay, bandwidth, probability of packet loss, signal strength, etc., are vital in order to get enhanced performance in terms of PDR and end-to-end delay. It facilitates reduced number of route reconfigurations in the network in case of link or route failure.
	\par Large number of approaches with QoS support have been published for mesh-based protocol. Due to highly dynamic topology of network, providing QoS support is very difficult. A QoS modification can be executed at different layers according to the application requirements. At MAC layer, we can determine delay and packet loss ratio. Similarly, we can obtain received signal strength, bit error rate and transmission quality at physical layer.
	\par Real-time video streaming requires a minimum bandwidth to communicate, so we can add QoS parameter to enhance transmission quality and bandwidth assurance.
	\item[(v)] \textit{\textbf{Secure Multicasting:}} In MANETs, secure networking has become a subject of great concern to researchers because wireless networks are more prone to passive and active attacks. In multicast scenario, security is more sophisticated due to number of receivers attached to the network. A single attacker node can degrade the performance of entire network.
	\par The multicast routing protocol should be efficient to provide protection from denial of service attacks, misbehaving nodes, unauthorized access to data, etc. We can make MANETs secure from unauthorized access of data by applying encryption mechanisms with group key management. Although, we have to mitigate excessive overhead that would be generated due to cryptography techniques. In addition, mobile nodes run on limited energy resources and have low computing power thus, applying such complex techniques would drain off available resources. 
	\par In MANETs, security is receiving additional attention due to infrastructure less network, no central administration, dynamic topology, etc. Several solutions have been proposed for security in MANETs, but not much light-weighted mechanisms. Mainly, approaches are based on delay calculation, behavior analysis, trust based and geo-casting. Geo-casting is used for calculating node position in battlefield to check authenticity of node because single node can be caught and made malicious. Thus, we require a flexible and high security mechanism that can adapt in all conditions discussed above. All these conditions are difficult to implement in order to secure multicast routing protocol for multicast communication.
\end{itemize}
\section{Future Directions for Multicast Routing Protocol in MANETs}
\label{future}
%We have confidence 
We believe that investigations for employing multicast communications in MANETs is still in its initial stage. It is mandatory to improve the protocol to be more robust and reliable in case of link failure or dynamic changes in topology of the network. We have defined some open issues like QoS guarantee, reliability, power efficiency and security provisioning that are essential for an effective routing protocol. It is difficult to design an MRP that successfully takes all these issues into consideration. We can also refine link stability parameters, that are defined in the last section, to improve the protocol performance. We can combine different routing modification mechanisms to make more resilient protocol.
\par On the basis of our understanding of multimedia data transfer over multicast MANETs, following may be taken up as an extension in future being research challenges:
\begin{enumerate}
	\item We will work to consolidate a system using our approaches to establish uninterrupted communication in a real-time application.
	\item Develop a robust QoS-aware multicast routing protocol that can reliably transmit video streaming data over flying adhoc networks~\cite{fanet}. Our protocol should be reliable with all the constraints in FANETs such as scalability and high bandwidth.
\end{enumerate}
\section{Conclusion}
\label{conclude}
This paper discusses the state of the art research in mesh based multicast routing protocols in MANETs. From discussions as presented earlier, it can be inferred that selecting QoS metric for the specific problem domain is significant especially in MRP. A suitable QoS metric is useful in assessing "goodness" of a routing solution as per requisite performance. Various enhancements in ODMRP have been discussed on the basis of routing modifications and Quality of Services parameters. Protocols have been categorized on the basis of type of modifications to achieve better throughput in terms of packet delivery ratio, end-to-end delay, control overhead and packet loss ratio. A critical review of existing multicast routing protocols have been presented; and each protocol is discussed with its advantages and limitations. Issues regarding multicast routing protocols in MANETs are discussed in this paper.

% In the next chapter, we shall be presenting our proposal on link stable multicast route protocol. The proposed solution selects a reliable route for multicast communication in MANETs. Subsequently, a novel mobility prediction method is applied on a mobile node to predict the node movement for uninterrupted transmission.
%This paper has opened up new avenues for research in mesh based multicast routing protocols in MANETs. It offers a view of selecting QoS metric for the problem definitions get to be solved. We have discussed various enhancements in ODMRP on the basis of Routing Modifications and Quality of Services parameters. We have categorized protocols on the basis of their type of modifications to achieve better throughput in terms of Packet Delivery Ratio, End-to-End Delay, Control Overhead and Packet Loss Ratio. Every aspect has been critically surveyed with its advantages and limitations for respective protocol. We have also discussed various issues regarding multicast routing protocols in MANETs.
\balance

\par The author(s) declare(s) that there is no conflict of interest regarding the publication of this paper.

\end{document}